\documentclass[aps,prb,twocolumn,showpacs,superscriptaddress]{revtex4}
\usepackage{graphicx} 
\newcommand{\be}{\begin{equation}}
\newcommand{\ee}{\end{equation}}
\newcommand{\bea}{\begin{eqnarray}}
\newcommand{\eea}{\end{eqnarray}}

\begin{document}

\title{Synthesis, Structure and Properties of Tetragonal
Sr$_{2}$$M_{3}$As$_{2}$O$_{2}$ \\($M_3$ = Mn$_3$,  Mn$_2$Cu and MnZn$_2$) Compounds Containing\\ Alternating CuO$_2$-Type and FeAs-Type Layers}

\author{R. Nath}
\altaffiliation{Present address: Indian Institute of Science Education and Research, Thiruvananthapuram-695016, Kerala, India}
\affiliation{Ames Laboratory and
Department of Physics and Astronomy, Iowa State University, Ames, Iowa 50011, USA}
\author{V. O. Garlea}
\affiliation{Neutron Scattering Sciences Division, Oak Ridge National Laboratory, Oak Ridge, Tennessee 37831, USA}
\author{A. I. Goldman}
\author{D. C. Johnston}
\affiliation{Ames Laboratory and
Department of Physics and Astronomy, Iowa State University, Ames, Iowa 50011, USA}

\date{\today}

\begin{abstract}
Polycrystalline samples of Sr$_{2}$Mn$_{2}$CuAs$_{2}$O$_{2}$, 
Sr$_{2}$Mn$_{3}$As$_{2}$O$_{2}$, and Sr$_{2}$Zn$_{2}$MnAs$_{2}$O$_{2}$ were synthesized. Their temperature- and applied magnetic field-dependent structural, transport, thermal, and magnetic properties were characterized by means of x-ray and neutron diffraction, electrical resistivity $\rho$, heat capacity, magnetization and magnetic susceptibility measurements.  These compounds have a body-centered-tetragonal crystal structure (space group \emph{I}4/\emph{mmm}) that consists of $M$O$_2$ ($M$ = Zn and/or Mn) oxide layers similar to the CuO$_2$ layers in high superconducting transition temperature $T_{\rm c}$ cuprate superconductors, and intermetallic $M$As ($M$ = Cu and/or Mn) layers similar to the FeAs layers in high-$T_{c}$ pnictides.  These two types of layers alternate along the crystallographic $c$-axis and are separated by Sr atoms.  The site occupancies of Mn, Cu and Zn were studied using Rietveld refinements of x-ray and neutron powder diffraction data. The temperature dependences of $\rho$ suggest metallic character for Sr$_{2}$Mn$_{2}$CuAs$_{2}$O$_{2}$ and semiconducting character for Sr$_{2}$Mn$_{3}$As$_{2}$O$_{2}$ and Sr$_{2}$Zn$_{2}$MnAs$_{2}$O$_{2}$.  Sr$_{2}$Mn$_{2}$CuAs$_{2}$O$_{2}$ is inferred to be a ferrimagnet with a Curie temperature $T_{\rm C} = 95(1)$~K\@.  Remarkably, we find that the magnetic ground state structure changes from a G-type antiferromagnetic structure in Sr$_{2}$Mn$_{3}$As$_{2}$O$_{2}$ to an A-type ferrimagnetic structure in Sr$_{2}$Mn$_{2}$CuAs$_{2}$O$_{2}$ in which the Mn ions in each layer are ferromagnetically aligned, but are antiferromagnetically aligned between layers.
\end{abstract}


\maketitle

\section{Introduction}

Superconductivity research was reinvigorated by the discovery of high temperature
superconductivity in layered
cuprates in 1986.\cite{bednorz1986} The highest superconducting transition temperature $T_{c}$
of $164$~K reported up to now for any material was achieved in
HgBa$_{2}$Ca$_{2}$Cu$_{3}$O$_{8+\delta}$ under pressure.\cite{schilling1993,
gao1994}
Recently a series of layered pnictide compounds $R$FeAsO$_{1-x}$F$_{x}$
($R$ = La, Ce, Pr, Nd, Sm, and Gd) \cite{ren2008, kamihara2008,
xhchen2008,
ren2008a, kamihara2006} and $A_{1-x}$K$_x$Fe$_2$As$_2$ ($A$ = Ba, Sr,
Ca, and Eu) (Refs. \onlinecite{rotter2008a, gfchen2008a, jeevan2008a,
sasmal2008,
wu2008a}) and other pnictide families were discovered where $T_{c}$ ranges up to
$56$~K\@.  Both the cuprate and iron pnictide families of high-$T_{\rm c}$ compounds contain square lattice layers, of Cu and Fe, respectively.
A principal difference between the cuprate and pnictide superconductors is in the detailed nature of the transition metal layers. In the cuprates, the oxygen atoms in the CuO$_2$ layers are situated directly between nearest-neighbor Cu atoms, the bonding in the plane is primarily Cu-O bonding, and the states at the Fermi energy $E_{\rm F}$ contain a large O 2$p$ component.  The high $T_{c}$ in the cuprates is believed to be intimately related to the crystal geometry.\cite{attfield1998, lin1995, varela2002}  In the FeAs-type compounds, on the other hand, the As atoms are arranged in layers on either side of the Fe square lattice layers, resulting in tetrahedral coordination of the Fe atoms by As.  Strong direct Fe-Fe interactions cause the electron states at $E_{\rm F}$ to be dominated by states derived from the Fe $d$-orbitals.  

A real challenge is to search for new materials with even higher $T_{c}$ values.  As is evident from the previous studies, $T_{c}$ is intimately related to the arrangement, separation, and atomic constituents of the transition metal layers.  Therefore it may be possible to achieve an enhanced
$T_{c}$ by tuning certain structural  and/or chemical parameters.  Recently 
Volkova\cite{volkova2008} and Ozawa \emph{et al.}\cite{ozawa2008} reported comparative studies of the structural properties of different layered compounds and suggested that
$A_{2}$Mn$_{3}$$Pn_{2}$O$_{2}$ ($A$ = Ba, Sr; pnictogen $Pn$ = As, Sb) type compounds might serve as parent compounds for high-$T_{c}$ superconductivity. The structure of this class of compounds contains two
different types of layers --- an $M$O$_2$ layer similar to the CuO$_2$ layers in the high-$T_{c}$ cuprates
and an $M^\prime$As layer similar to the FeAs layers in the pnictide
superconductors.\cite{brechtel1979,brock1996}  Magnetic and structural properties of some of these compounds have been studied via magnetization\cite{brock1996a,stetson1991} and neutron diffraction\cite{brock1996} measurements.  There has also been much recent experimental interest in mixed layered chalcogenide
oxide\cite{park1993} and oxysulfide\cite{zhu1997} compounds as potential condidates for high-$T_{c}$ compounds and/or for other interesting magnetic and electronic properties.\cite{kaczorowski1994}

\begin{figure}
\includegraphics [width=3.3in]{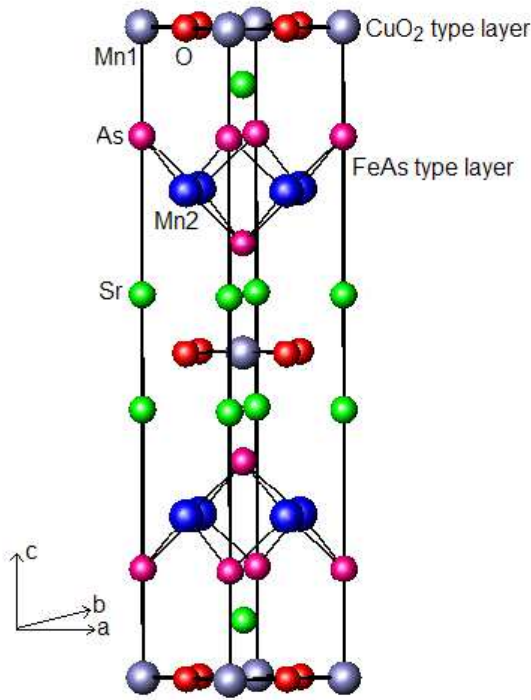}
\caption{\label{structure} (Color online) Crystal structure of
Sr$_2$Mn$_{3}$As$_{2}$O$_{2}$ showing alternating CuO$_{2}$-type and
FeAs-type layers along the $c$-axis.  These layers are separated by
layers of Sr atoms.}
\end{figure}

The $A_{2}$Mn$_{3}$$Pn_{2}$O$_{2}$ compounds crystallize in the body-centered-tetragonal tetragonal space group $I$4$/mmm$. As shown in Fig.~\ref{structure},
the parent compound Sr$_{2}$Mn$_{3}$As$_{2}$O$_{2}$ has two
inequivalent Mn sites. The Mn1 atoms are fourfold coordinated by O in a plane to
form a CuO$_2$-type square lattice as in the high-$T_{c}$ cuprates.
The Mn2 atoms are fourfold coordinated by As to form FeAs-type 
layers similar to the FeAs layers in the high-$T_{\rm c}$ pnictide superconductors.  The 
CuO$_2$-type and FeAs-type layers alternate along the crystallographic $c$~axis and are separated by Sr layers.  Each type of Mn atom has four nearest-neighbor Mn atoms within each plane, and four nearest-neighbors in each of the two adjacent planes.  An interesting and potentially important aspect of this arrangement of the Mn atoms in adjacent layers is that the interactions between Mn1 and Mn2 are  geometrically frustrated for antiferromagnetic ordering.  Previous neutron diffraction measurements showed that there are no structural transitions down to 4~K that could serve to relieve the frustration.\cite{brock1996, ozawa2001}

The syntheses and magnetic measurements of Sr$_{2}$Zn$_{2}$MnAs$_{2}$O$_{2}$ and Sr$_{2}$Mn$_{3}$As$_{2}$O$_{2}$ have been previously reported.\cite{brock1996, brock1996a, ozawa1998,ozawa2001} For Sr$_{2}$Mn$_{3}$As$_{2}$O$_{2}$,
magnetic neutron diffraction measurements indicate the presence of two 
distinct types of magnetic sublattice.\cite{brock1996} The Mn2 sublattice undergoes long-range antiferromagnetic (AF) ordering at a N\'eel temperature $T_{\rm N} = 340$~K in a G-type AF structure with an ordered moment of 3.4~$\mu_{\rm B}$/Mn at 4~K, where $\mu_{\rm B}$ is the Bohr magneton.\cite{brock1996}  On the other hand, the Mn1 sublattice does not undergo long-range ordering down to 4~K but instead shows very weak magnetic 
reflections below 75~K where the most prominent reflection has a Warren line shape, which is indicative of two-dimensional short-range order.\cite{brock1996}  Zero-field-cooled (ZFC) and field-cooled (FC) $\chi(T)$ data show a broad but weak maximum at 65~K and a bifurcation at 51~K\@.\cite{brock1996a}   For Sr$_{2}$MnZn$_{2}$As$_{2}$O$_{2}$, 
ZFC and FC susceptibility data show a weak maximum at 65~K and a splitting around 30~K\@.\cite{ozawa2001}  A Curie-Weiss fit to $\chi(T)$ at high $T$ yields a positive Weiss temperature $\theta=43$~K suggesting that the dominant interaction is ferromagnetic.  However, neutron diffraction  experiments show no evidence for either short-range or long-range order down to 4~K\@.\cite{ozawa2001} 

Herein, we report the synthesis and detailed characterization of the Sr$_{2}$Mn$_{3}$As$_{2}$O$_{2}$-type compound Sr$_{2}$Mn$_{2}$CuAs$_{2}$O$_{2}$.  This compound has not been synthesized or studied before to our knowledge, although Cu-containing materials were previously suggested.\cite{BrockJuly1995}  The goal was to synthesize a parent compound with CuO$_2$ layers as in the high-T$_{\rm c}$ layered cuprates, alternating with MnAs layers as in the FeAs-type materials, which upon doping might become a high-$T_{\rm c}$ superconductor.  However, we found that although the compound with the desired composition does form, the Cu atoms do not go into the CuO$_2$-type layers, but rather statistically occupy approximately half of the transition metal sites in the FeAs-type layers.  We have characterized this material in detail as a function of temperature $T$ and/or applied magnetic field $H$ by means of x-ray and neutron diffraction, electrical resistivity $\rho$, heat capacity $C_{\rm p}$, magnetization $M$ and magnetic susceptibility $\chi$ measurements.  For comparison, we also carried out detailed structure and property measurements on the above previously reported Sr$_{2}$Zn$_{2}$MnAs$_{2}$O$_{2}$ and Sr$_{2}$Mn$_{3}$As$_{2}$O$_{2}$ compounds.

The remainder of the paper is organized as follows.  The synthesis and measurement details are given in Sec.~\ref{SecExpDet}.  The results of our x-ray and neutron structure analyses, magnetization and magnetic susceptibility, and heat capacity measurements are given in Sec.~\ref{SecResults}.  A discussion of our results is given in Sec.~\ref{SecDiscuss}.  A summary of our results and conclusions is given in Sec.~\ref{SecSummary}.

\section{\label{SecExpDet} Experimental Details}

\subsection{Sample Preparation}

Polycrystalline samples of Sr$_{2}$Mn$_{2}$CuAs$_{2}$O$_2$,
Sr$_{2}$Zn$_{2}$MnAs$_{2}$O$_2$, and Sr$_{2}$Mn$_{3}$As$_{2}$O$_{2}$
were prepared by solid state reaction techniques using
SrO ($99.9$\% pure), Mn ($99.99$\% pure, Alfa-Aesar), Cu ($99.99$\% pure, Fisher), Zn
($99.99$\% pure, Alfa-Aesar), and
As ($99.9$\% pure, Alfa-Aesar) as starting materials. The SrO was prepared by heating SrCO$_3$ (99.99\% pure, Aldrich) at 1300~$^\circ$C in air for 12 h and cooling under vacuum.  The stoichiometric mixtures of the starting materials were placed in an Al$_2$O$_3$
crucible that was then sealed inside an evacuated quartz tube.  The
samples were first heated to 610~$^{\circ }$C at a rate of
$80$~$^{\circ }$C/h, held there for 10~h and then heated to
$980$~$^{\circ }$C and held there for 20~h. The samples were
then progressively fired at 980~$^{\circ }$C and 1000~$^{\circ
}$C for 30~h, each followed by one intermediate grinding and
pelletization.  All the sample handling was carried out inside a He-filled glove box.

\subsection{X-ray, Magnetic Susceptibility, Heat Capacity, and Electrical Resistivity Characterization}

The samples were characterized using a Rigaku Geigerflex powder
diffractometer with a Cu target ($\lambda _{\rm ave}=1.54182$~\AA).
The magnetization $M(H,T)$ and magnetic susceptibility $\chi(T) \equiv M(H,T)/H$ were measured
in the temperature $T$ range 1.8~K $\leq T \leq 350$~K in
applied fields up to 5.5~T\@. Zero-field-cooled (ZFC) and
field-cooled (FC) magnetic susceptibilities were also measured as a function
of $T$ at $H = 100$~Oe.  The magnetic measurements were carried out using a commercial
(Quantum Design) SQUID (superconducting quantum interference device) magnetometer.  The DC resistivity $\rho(T)$ was measured with a standard four-probe technique using a current of 5~mA, and heat capacity $C_{\rm p}(T)$ was measured on samples of mass $\sim 5$~mg. The $\rho(T)$ and $C_{\rm p}(T)$ measurements were performed on pieces of sintered pellets using a Quantum Design Physical Property Measurement System (PPMS).

\subsection{Neutron Diffraction Measurements}

Neutron powder diffraction measurements on both
Sr$_{2}$Mn$_{3}$As$_{2}$O$_{2}$ and
Sr$_{2}$Mn$_{2}$CuAs$_{2}$O$_{2}$ were carried out at the HB2A
neutron powder diffractometer at the High Flux Isotope Reactor at
Oak Ridge National Laboratory, using a wavelength of $\lambda$ =
1.536~{\AA} provided by a vertically focusing Ge(115) monochromator.
For data collection, the detector array consisting of 44 $^3$He tubes
was scanned in two segments to cover the total 2$\theta$ range of
4$^{\circ}$ to 150$^{\circ}$, in steps of 0.05$^\circ$. Overlapping
detectors for a given step served to average the counting efficiency
of each detector. More details about the HB2A instrument and data
collection strategies can be found in Ref.~\onlinecite{HB2A}.
Measurements were made on approximately 5~g of sample held in
a cylindrical vanadium container in a top-loading closed cycle
refrigerator (4--300~K).

For the Sr$_{2}$Mn$_{3}$As$_{2}$O$_{2}$ sample, powder patterns were
collected at $T = 375$~K (above $T_{\rm N}$ as reported by Brock et
al.~\cite{brock1996}), 300~K, 150~K, 75~K and at 4~K, below the temperature
where new magnetic reflections have been reported.\cite{brock1996}  For
Sr$_{2}$Mn$_{2}$CuAs$_{2}$O$_{2}$, powder patterns were collected at
$T = 375$~K, 300~K, 150~K, 60~K and at 4~K\@.  Rietveld refinements were
performed using the FULLPROF program.\cite{fullprof} Although both
samples were relatively phase pure, small amounts of impurity phases of MnO 
($\approx 2$--3\%) and a second, unidentified, phase were present.  MnO
orders antiferromagnetically below the N\'eel temperature $T_{\rm N} \approx 118$~K,\cite{Morosin1970} and the
associated magnetic peaks were identified during the refinements.

\section{\label{SecResults} RESULTS}

\subsection{Structure}

\begin{figure}
\includegraphics [width=3.5in]{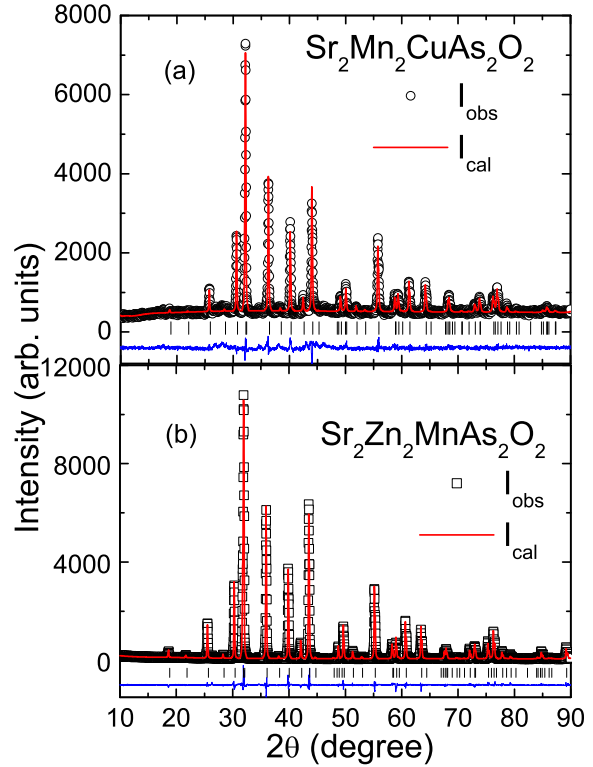}
\caption{\label{xrd}
(Color online) X-ray powder diffraction patterns (open circles) at
room temperature for (a) Sr$_{2}$Mn$_{2}$CuAs$_{2}$O$_2$ and (b)
Sr$_{2}$Zn$_{2}$MnAs$_{2}$O$_2$. The solid lines are
Rietveld refinement fits with $I$4$/mmm$ space group. For
Sr$_{2}$Mn$_{2}$CuAs$_{2}$O$_2$ the fit shown is for model (ii) discussed in the text.}
\end{figure}

\begin{table}
\caption{\label{Occupation} Site occupancy models and quality of fit parameters $R$ for
Sr$_{2}$Mn$_{2}$CuAs$_{2}$O$_2$ from Rietveld refinements of powder XRD data at room
temperature.  The 2($a$) site is the transition metal site in the CuO$_2$-type layer, whereas the 4($d$) site is the transition metal site in the FeAs-type layer.}
\begin{ruledtabular}
\begin{tabular}{cccccc}
Model & $2(a)$ site & $4(d)$ site  & $R$ (\%)\\ \hline
 (i) & Cu & Mn & 16 \\
(ii)  & Mn & Mn and Cu & 18 \\
(iii)  & Mn and Cu & Mn and Cu & 13.7 \\
\end{tabular}
\end{ruledtabular}
\end{table}

The powder x-ray diffraction patterns indicated almost single phase
samples of Sr$_{2}$Mn$_{2}$CuAs$_{2}$O$_2$ and
Sr$_{2}$Zn$_{2}$MnAs$_{2}$O$_2$ while in Sr$_{2}$Mn$_{3}$As$_{2}$O$_{2}$, 
about 2.5$\%$ MnO impurity phase was detected. Rietveld
refinements of the data were carried out using the GSAS
package.\cite{larson2000} The initial crystallographic parameters were
taken from the previous reports on Sr$_{2}$Mn$_{3}$As$_{2}$O$_2$
(Refs.~\onlinecite{brechtel1979} and \onlinecite{brock1996}) and Sr$_{2}$Zn$_{2}$MnAs$_{2}$O$_2$
(Ref.~\onlinecite{ozawa2001}). In addition to the
other parameters, the transition metal site [2$(a)$ (0 0 0) and
4$(d)$ (0 0.5 0.25)] occupancies were also refined.  We 
considered three models for Sr$_{2}$Mn$_{2}$CuAs$_{2}$O$_2$ as listed in Table~\ref{Occupation}. In
model~(ii), the 4$(d)$ site is occupied by 50\% Mn and 50\% Cu and in
model~(iii), the 2$(a)$ site is occupied by 50 \% Mn and 50 \% Cu
and the 4$(d)$ site by 75 \% Mn and 25 \% Cu.  The relatively large $R$ value in each case is mainly due to
our inability to fit the background precisely. Since the $R$ values in the three cases
are nearly the same, we cannot draw any conclusions about the respective occupancies of the 2($a$) and 4($d$) sites.  Figure~\ref{xrd}(a) shows the Rietveld
refinement fit to the powder x-ray diffraction pattern obtained using model~(ii).

On the other hand, for Sr$_{2}$Zn$_{2}$MnAs$_{2}$O$_2$, using refinements similar to those above, Mn and Zn were found to only occupy the CuO$_2$-type and FeAs-type layers, respectively. Figure~\ref{xrd}(b) presents the Rietveld refinement for Sr$_{2}$Zn$_{2}$MnAs$_{2}$O$_2$, for which the $R$-factor is $R_{\rm wp} = 6$\%.  This result is very similar to the previous report on
Sr$_{2}$Zn$_{2}$MnAs$_{2}$O$_{2}$ where powder neutron diffraction
data indicated complete site occupancy of the oxide layer by Mn and
occupancy of the pnictide layer by 94\% Zn and 6\% Mn.\cite{ozawa2001} It
was also mentioned that the lack of full occupancy in the
pnictide layer by Zn might be due to the loss of Zn during synthesis
since a significant amount of Zn$_{2}$As$_{3}$ separated from the
starting materials by gas phase diffusion.  Moreover similar site occupancies were found for the Ba analogue 
Ba$_{2}$Zn$_{2}$MnAs$_{2}$O$_{2}$.\cite{ozawa1998}

\begin{table}
\caption{\label{Refinement} Structure parameters for
Sr$_{2}$Mn$_{2}$CuAs$_{2}$O$_2$ (space group $I$4$/mmm$, $Z = 2$~f.u./unit cell) refined from powder XRD data at room
temperature for model (ii).}
\begin{ruledtabular}
\begin{tabular}{cccccc}
Atom & Wyckoff & $x$ & $y$ & $z$ & $B$  \\ 
& site & & & & (\AA$^{2}$) \\\hline
 Sr & $4(e)$ & 0 & 0 & 0.4104(1) & 0.009(1) \\
 Mn1 & $2(a)$ & 0 & 0 & 0 & 0.023(3) \\
 Cu & $4(d)$ & 0 & 0.5 & 0.25 & 0.041(4) \\
 Mn2 & $4(d)$ & 0 & 0.5 & 0.25 & 0.000 \\
 As & $4(e)$ & 0 & 0 & 0.1675(2) & 0.018(2) \\
 O & $4(c)$ & 0 & 0.5 & 0 & 0.053(9) \\
\end{tabular}
\end{ruledtabular}
\end{table}

The lattice parameters obtained at room temperature are [$a=4.0833(2)$~\AA,
$c=18.5919(9)$~\AA], [$a=4.12757(6)$~\AA, $c=18.6941(4)$~\AA], 
and [$a=4.14160(7)$~\AA, $c=18.8177(4)$~\AA] for 
Sr$_{2}$Mn$_{2}$CuAs$_{2}$O$_2$, Sr$_{2}$Zn$_{2}$MnAs$_{2}$O$_2$, 
and Sr$_{2}$Mn$_{3}$As$_{2}$O$_{2}$, respectively.  The values for
Sr$_{2}$Zn$_{2}$MnAs$_{2}$O$_2$ and Sr$_{2}$Mn$_{3}$As$_{2}$O$_{2}$ 
are close to the previously reported
ones\cite{ozawa2001} while for the new compound Sr$_{2}$Mn$_{2}$CuAs$_{2}$O$_2$, the
values are slightly smaller than those reported for the other two compounds.  
Some parameters obtained from the Rietveld refinements on
Sr$_{2}$Mn$_{2}$CuAs$_{2}$O$_2$ are listed in Table~\ref{Refinement}.

\begin{table}
\caption{\label{compounds} Compositions synthesized during unsuccessful attempts to obtain single-phase compounds with the Sr$_{2}$Mn$_{3}$As$_{2}$O$_2$-type structure.}
\begin{ruledtabular}
\begin{tabular}{ll}
 Sr$_{2}$Fe$_{3}$As$_{2}$O$_2$ & Sr$_{2}$Zn$_{2}$CuAs$_{2}$O$_{2}$  \\
 Sr$_{2}$Fe$_{3}$P$_{2}$O$_{2}$ & Sr$_{2}$Mn$_{2}$CuSb$_{2}$O$_{2}$ \\
 Sr$_{2}$Fe$_{3}$Sb$_{2}$O$_{2}$ & Ca$_{2}$Mn$_{2}$CuAs$_{2}$O$_{2}$ \\
 Sr$_{2}$Fe$_{2}$CuAs$_{2}$O$_{2}$ & Sr$_{2}$MnCu$_{2}$As$_{2}$O$_{2}$ \\
 Sr$_{2}$Fe$_{2}$CuSb$_{2}$O$_{2}$ & Sr$_{2}$Zn$_{2}$FeAs$_{2}$O$_{2}$ \\
 Ca$_{2}$Fe$_{2}$CuAs$_{2}$O$_{2}$ & Sr$_{2}$Mn$_{2}$FeAs$_{2}$O$_{2}$ \\
 Sr$_{2}$Mn$_{2}$CuP$_{2}$O$_{2}$ & Sr$_{2}$Ni$_{2}$ZnAs$_{2}$O$_{2}$ \\
\end{tabular}
\end{ruledtabular}
\end{table}

Our main goal was to achieve a material with Fe occupying the MnAs 
layer and Cu occupying the MnO$_2$ layer.  With this objective in mind, in addition to the three compounds discussed in this paper that were obtained in nearly single-phase form, we attempted to synthesize fourteen other compounds with the Sr$_{2}$Mn$_{3}$As$_{2}$O$_2$-type structure that are listed in Table~\ref{compounds}. We did not obtain single-phase materials with these compositions and these are therefore not discussed further.

\subsection{Electrical Resistivity Measurements}

\begin{figure}
\includegraphics [width=3in]{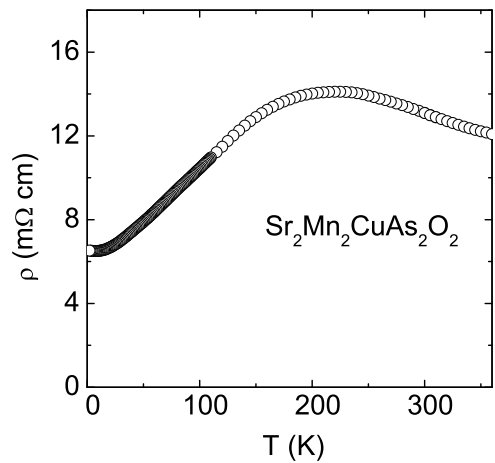}
\caption{\label{resist1}
DC electrical resistivity $\rho$ versus temperature $T$ of
Sr$_{2}$Mn$_{2}$CuAs$_{2}$O$_{2}$.}
\end{figure}

\begin{figure}
\includegraphics [width=3in]{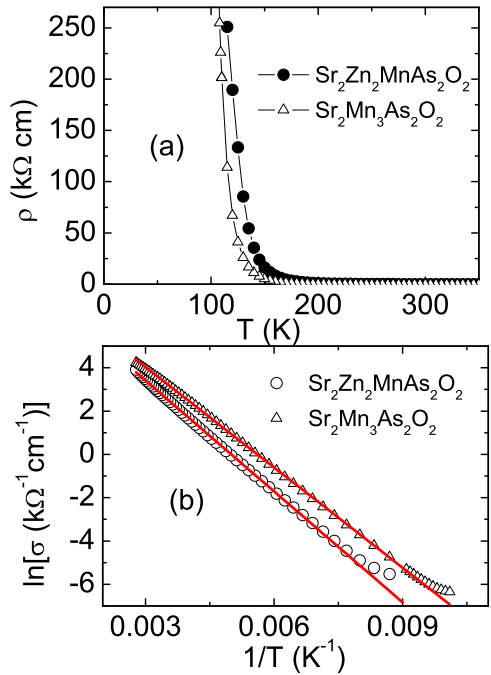}
\caption{\label{resist2}
(Color online) (a) DC electrical resistivity $\rho$ of Sr$_{2}$Mn$_{3}$As$_{2}$O$_{2}$ and
Sr$_{2}$Zn$_{2}$MnAs$_{2}$O$_{2}$ versus temperature $T$. (b)
ln($\sigma$) versus $1/T$.  The red straight lines are linear fits to
the data between 150 and 300~K\@.}
\end{figure}

Figure~\ref{resist1} shows $\rho(T)$ for Sr$_{2}$Mn$_{2}$CuAs$_{2}$O$_{2}$, and
Fig.~\ref{resist2}(a) shows $\rho(T)$ for Sr$_{2}$Mn$_{3}$As$_{2}$O$_{2}$ and 
Sr$_{2}$Zn$_{2}$MnAs$_{2}$O$_{2}$. With decreasing $T$, $\rho(T)$ for Sr$_{2}$Mn$_{2}$CuAs$_{2}$O$_{2}$ first increases, then shows a broad maximum at about 200~K, and then decreases to a residual resistivity of about $6.5$~m$\Omega$~cm at 2~K\@.  The decrease in $\rho(T)$ towards a constant value for $T \to 0$~K
suggests a metallic ground state. The negative coefficient of resistivity at high $T$ and an overall
magnitude of resistivity higher than expected for a metal may be associated with trace amounts of high-resistivity impurities in the grain boundaries of the polycrystalline sample as often occurs in oxides.

In contrast, for Sr$_{2}$Mn$_{3}$As$_{2}$O$_{2}$ and 
Sr$_{2}$Zn$_{2}$MnAs$_{2}$O$_{2}$, $\rho$ increases
monotonically with decreasing $T$ towards large values, pointing
towards an insulating ground state of the compounds.  Data below
$115$~K were not obtained for these compounds since the resistances of the samples below
this temperature exceeded the measurement limit of the equipment.
Figure~\ref{resist2}(b) shows a plot of ln$\sigma$ versus $1/T$ 
where $\sigma=1/\rho$ is the conductivity.  We
fitted the data between 120~K and 300~K and between 150 and 300~K by the expression $\ln\sigma =
A + \Delta/(k_{\rm B}T)$ as shown in Fig.~\ref{resist2}(b) where $A$ is
a constant and $\Delta$ is the activation energy, yielding $\Delta =
133(4)$~meV and 147(5)~meV for Sr$_{2}$Mn$_{3}$As$_{2}$O$_{2}$ and 
Sr$_{2}$Zn$_{2}$MnAs$_{2}$O$_{2}$, respectively.  We infer that 
Sr$_{2}$Mn$_{3}$As$_{2}$O$_{2}$ and Sr$_{2}$Zn$_{2}$MnAs$_{2}$O$_{2}$ 
are narrow band gap semiconductors.

\subsection{Magnetization and Magnetic Susceptibility Measurements}

\subsubsection{\rm Sr$_{2}$Zn$_{2}$MnAs$_{2}$O$_2$}

\begin{figure}
\includegraphics [width=3in]{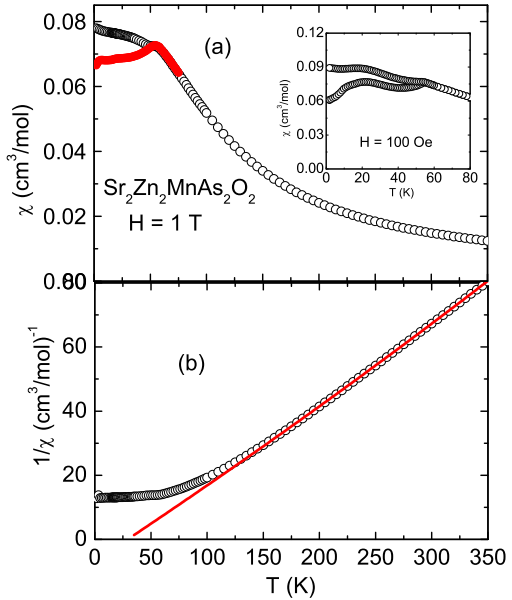}
\caption{\label{susc2} (Color online) (a) ZFC (red, lower data set) and FC (black, upper data set) magnetic
susceptibility $\chi$ of
Sr$_{2}$Zn$_{2}$MnAs$_{2}$O$_2$ versus temperature $T$ in magnetic
field $H = 1$~T\@. Inset: ZFC and FC susceptibilities versus $T$
at $H = 100$~Oe. (b) Inverse magnetic susceptibility
$\chi^{-1}$ versus $T$.  The solid line is a Curie-Weiss fit to the 
high temperature data between 175 and 350~K\@.}
\end{figure}

The magnetic susceptibility $\chi \equiv M/H$ as a function of
temperature $T$ is
shown in Fig.~\ref{susc2}(a) for Sr$_{2}$Zn$_{2}$MnAs$_{2}$O$_2$ measured 
at an applied magnetic field $H = 1$~T\@.
$\chi$ increases with decreasing $T$, suggesting local moment
magnetism.  The ZFC and FC susceptibilities show a bifurcation below 
$\approx 50$~K, as also seen for $H = 100$~Oe in the inset of Fig.~\ref{susc2}(a), indicating the occurrence of magnetic ordering of some type below that temperature.

To fit the molar magnetic susceptibility data of a local moment
system in the paramagnetic state, one often uses the expression
\begin{equation}
\chi=\chi_{0}+\frac{C}{T-\theta},
\label{Curie}
\end{equation}
where $\chi_{0}$ is the temperature-independent contribution.
The second term is the Curie-Weiss law with Curie constant
$C=N_{\rm A}\mu_{\rm eff}^{2}/(3k_{\rm B})$ and Weiss temperature
$\theta$, where $N_{\rm A}$ is Avogadro's number, $\mu_{\rm eff}$ is
the effective magnetic moment per formula unit, and $k_{\rm B}$ is
Boltzmann's constant.  For an insulator, one has
\begin{equation}
\chi_0 = \chi_{\rm core} + \chi_{\rm VV}
\label{chi_0}
\end{equation}
where $\chi_{\rm core}$ is the core diamagnetism and $\chi_{\rm VV}$
is the Van Vleck paramagnetism. $\chi_{\rm core}$ can be calculated
assuming an ionic model for individual atoms in the oxidation states
Sr$^{2+}$, Mn$^{2+}$, Cu$^{2+}$, Zn$^{2+}$, As$^{3-}$, and O$^{2-
}$,\cite{Magnetochemistry} yielding $\chi_{\rm core}= -1.6
\times 10^{-4}$~cm$^{3}$/mol.  The value of $\chi_{\rm VV}$ is not
easily accessible for a given compound, but when calculated it is
often found to have about the same magnitude as $\chi_{\rm core}$,
but with a positive instead of negative sign, yielding a net small
and often negligible value of $\chi_0$ in local moment systems.

Fitting $\chi(T)$ by Eq.~(\ref{Curie}) in the high-$T$ region 175--350~K yielded value of $\chi_0 = -6.8(5) \times 10^{-4}~{\rm cm^3/mol}$ that is far more negative than the above $\chi_{\rm core}$ value and hence is unphysical.  Therefore, we fitted the data  by fixing $\chi_{0}$ to either $\chi_{\rm core}$ or zero in Eq.~(\ref{Curie}), over different temperature ranges from 175~K to 350~K\@. Setting $\chi_{0}= \chi_{\rm core} = -1.6 \times 10^{-4}~{\rm cm^3/mol}$ yielded $C = 3.92(2)$~cm$^3$~K/mol and $\theta = 39(1)$~K, whereas setting $\chi_0 = 0$ yielded  $C = 3.84(3)$~cm$^3$~K/mol and $\theta = 41.6(16)$~K\@.  Taking into account both values of $C$ yields an effective moment $\mu_{\rm eff} = 5.57(5)~\mu_{\rm B}$/f.u.  These parameters are in
agreement with the previously reported values.\cite{ozawa2001}  The positive value of $\theta$ indicates that the dominant interactions in the compound are ferromagnetic.  The formal oxidation state of the Mn ions is +2, corresponding to a $d^5$ electronic configuration.  The observed $\mu_{\rm eff}$ is somewhat smaller than the value $g\sqrt{S(S+1)}\mu_{\rm B} = 5.92~\mu_{\rm B}$ expected for high-spin ($S = 5/2$) Mn$^{+2}$ with $g$-factor $g = 2$, possibly due to hybridization effects as apparently occurs in BaMn$_2$As$_2$.\cite{an2009}  For $S= 2$ with $g = 2$, one would instead obtain $\mu_{\rm eff}= 4.90~\mu_{\rm B}$, significantly smaller than the observed value.

\begin{figure}
\includegraphics [width=3.3in]{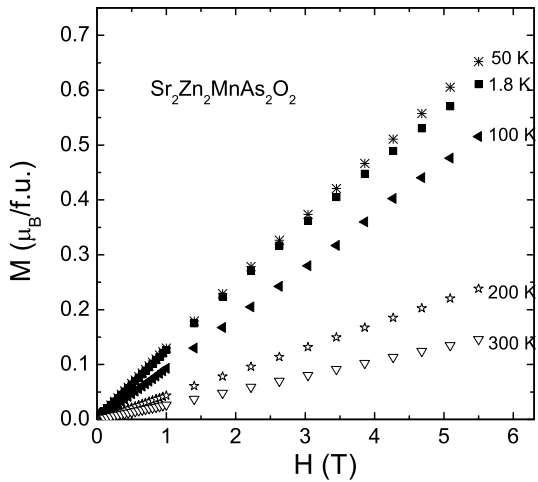}
\caption{\label{MH2} Magnetization $M$ versus applied magnetic field $H$ isotherms for
Sr$_{2}$Zn$_{2}$MnAs$_{2}$O$_2$ at the indicated temperatures.}
\end{figure}

Figure~\ref{MH2} shows $M(H)$ isotherms at different temperatures
between 1.8~K to 300~K for $H$ up to 5.5~T\@.  One sees that $M$ is
nearly proportional to $H$ at all temperatures down to $1.8$~K\@.  This
observation, and the bifurcation between the FC and ZFC
susceptibility data below $\approx 50$~K in the inset of Fig.~\ref{susc2}(a), are
consistent with the conclusion in Ref.~\onlinecite{ozawa2001} that 
spin-glass ordering occurs below $\sim 50$~K\@.

\subsubsection{\rm Sr$_{2}$Mn$_{3}$As$_{2}$O$_2$}

\begin{figure}
\includegraphics [width=3.3in]{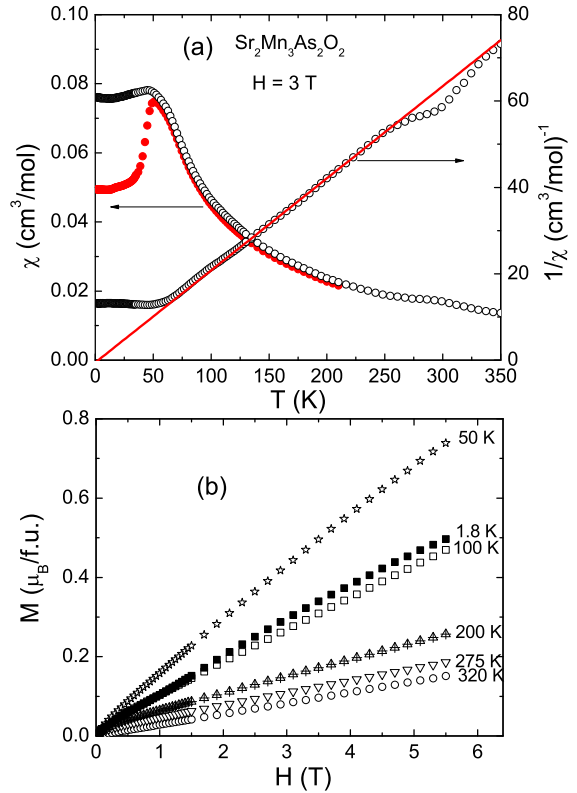}
\caption{\label{susc3} (Color online) (a) Zero-field-cooled ZFC (lower red data set) and field-cooled FC (upper black data set) magnetic susceptibility $\chi$ (left-hand axis) and $\chi^{-1}$ (right-hand axis) of Sr$_{2}$Mn$_{3}$As$_{2}$O$_2$ versus temperature $T$ in magnetic field $H = 3$~T\@. The data are (approximately) corrected for the contribution of $\approx 0.79$~mol\% of ferromagnetic MnAs impurity phase (see text).  The solid line is a Curie-Weiss fit of the data from 130~K to 250~K\@. (b) Magnetization $M$ versus applied magnetic field $H$ isotherms measured at the indicated temperatures.}
\end{figure}

Figure~\ref{susc3}(a) shows the ZFC and FC $\chi(T)$ of Sr$_{2}$Mn$_{3}$As$_{2}$O$_2$ 
measured at 3~T. A small step in $\chi(T)$ was observed at $T\simeq310$~K, even after correcting (see below) for a small amount of MnAs impurity which is known to have a first order ferromagnetic transition at 318~K.\cite{bean1962}  The ZFC and FC $\chi(T)$ data show a 
significant splitting at 50~K suggesting a possible spin-glass transition, a two-dimensional magnetic ordering transition,\cite{brock1996} or in view of our magnetic neutron diffraction data below, possibly a three-dimensional magnetic transition.  

$M(H)$ isotherms were measured at 
different temperatures to test for the presence of MnAs ferromagnetic impurity phase in the 
sample and are shown in Fig.~\ref{susc3}(b). At 320~K, $M$ is proportional to $H$ over the whole field range. Below $\sim 300$~K, nonlinearity was observed in the $M(H)$ curves below 0.5~T suggesting a small ferromagnetic MnAs impurity contribution to the magnetization which saturates by about 1~T\@. We did not, however, observe any peaks in our x-ray diffraction measurements corresponding to the fingerprint of MnAs, indicating that the MnAs impurity concentration is less than 2\% of the sample.  In order to quantitatively estimate the MnAs impurity concentration, we fitted the $M(H)$ isotherms for $H =  2.5$ to 5.5~T by the linear relation $M(H,T) = M_{\rm s}(T) + \chi(T)H$, where $M_{\rm s}(T)$ is the saturation magnetization of the FeAs ferromagnetic impurity phase and $\chi(T)$ is the intrinsic magnetic susceptibility of the sample.  
At 100~K, the value of $M_{s}=2.7\times 10^{-2}~\mu_{\rm B}$/f.u. corresponds to 
about 0.79~mol\% MnAs impurities [$M_{s}=3.40(3)~\mu_{\rm B}$/Mn for MnAs at 
$T$=0~K].\cite{haneda1977}  Our $\chi(T)$ data in Fig.~\ref{susc3}(a) are corrected for this ferromagnetic impurity contribution.  We evidently could not completely correct for this ferromagnetic contribution 
since the corrected $\chi(T)$ still shows a small step at 300~K\@.  This step is more 
pronounced in the $1/\chi$ plot. Our observations are in good agreement with those  
reported previously.\cite{brock1996a} 

Using the ionic model discussed above, $\chi_{\rm core}$ was calculated to be $-1.68\times10^{-4}$~cm$^3$/mol for Sr$_{2}$Mn$_{3}$As$_{2}$O$_2$.  We fitted the $\chi(T)$ data from 130~K to 250~K in Fig.~\ref{susc3}(a) by Eq.~(\ref{Curie}). The parameters $\chi_{0}$, $C$, and $\theta$ were 
found to be $-2.78(4)\times10^{-3}$ cm$^3$/mol, 5.7(2)~cm$^3$~K/mol,
and $-17(3)$~K, respectively.  The fitted value of $\chi_{0}$ is more negative than $\chi_{\rm core}$, which is unphysical since there are likely no other diamagnetic contributions to $\chi$.  Therefore in the following we fitted the data from 130~K to 250~K by fixing $\chi_{0}$ to either zero  or $\chi_{\rm core}$ in Eq.~(\ref{Curie}).  Setting $\chi_0 = 0$ yielded $C = 4.69(2)$~cm$^3$~K/mol
and $\theta = 3(1)$~K, whereas setting $\chi_0 = \chi_{\rm core}$ yielded $C = 4.75(2)$~cm$^3$~K/mol
and $\theta = 1.5(7)$~K\@.  Taking both values of $C$ into consideration yields $\mu_{\rm eff} = 6.14(2)~\mu_{\rm B}$/f.u.  This value is much smaller than the value of 10.3~$\mu_{\rm B}$/f.u.\ expected for three paramagnetic Mn$^{+2}$ spins per formula unit with $g = 2$ and $S = 5/2$.  However, the MnAs sublattice within Sr$_{2}$Mn$_{3}$As$_{2}$O$_2$ undergoes long-range AF order at a N\'eel temperature $T_{\rm N} = 340$~K with a G-type antiferromagnetic structure,\cite{brock1996} so only one out of the three Mn ions per formula unit (in the MnO$_2$ layer) should contribute to the Curie-Weiss law at lower $T$.  Our value $\mu_{\rm eff} = 6.14(2)~\mu_{\rm B}$/f.u.\ deduced at 130~K to 250~K is indeed close to the value of $5.92~\mu_{\rm B}$ expected for one spin $S = 5/2$ per formula unit with $g = 2$.

\subsubsection{\rm Sr$_{2}$Mn$_{2}$CuAs$_{2}$O$_2$}

\subsubsection*{Magnetic Susceptibility Measurements}

\begin{figure}
\includegraphics [width=3.5in]{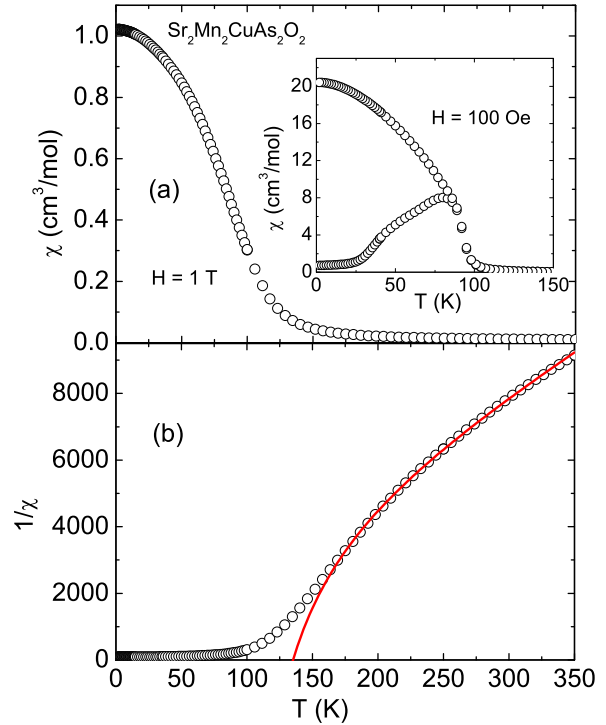}
\caption{\label{susc1} (Color online) (a) Magnetic susceptibility of
Sr$_{2}$Mn$_{2}$CuAs$_{2}$O$_2$ measured at $H =1$~T versus
temperature $T$. Inset shows ZFC and FC
susceptibility versus $T$ measured at $H = 100$~Oe. (b)
Inverse dimensionless volume susceptibility $1/\chi$ versus $T$.  The solid curve is a
fit to the $1/\chi$ data above 170~K by Eq.~(\ref{chi0a}).}
\end{figure}

The molar $\chi(T)$ of Sr$_{2}$Mn$_{2}$CuAs$_{2}$O$_2$ in $H = 1$~T is shown in
Fig.~\ref{susc1}(a). Below about 125~K, $\chi(T)$ increases rapidly,
suggesting the onset of ferrimagnetic/ferromagnetic ordering at a Curie temperature $T_{\rm C}\sim 100$~K in contrast to the spin-glass type transitions evidently observed above for Sr$_{2}$Zn$_{2}$MnAs$_{2}$O$_2$ and 
Sr$_{2}$Mn$_{3}$As$_{2}$O$_2$. The ZFC and FC susceptibilities in $H = 100$~Oe show a significant splitting at $90$~K as shown in the inset of Fig.~\ref{susc1}(a).  Such irreversibility is a characteristic
behavior of ferrimagnetic/ferromagnetic compounds.\cite{he2007} The magnetization isotherm data below confirm this interpretation.

Figure~\ref{susc1}(b) shows 1/$\chi$ versus $T$ where in this case $\chi$ is the dimensionless volume susceptibility, which is the molar susceptibility divided by the molar volume $V_{\rm M}$, both in units of cm$^3$/mol.  We plot here the inverse volume susceptibility in order to fit the data later by theory.  In the paramagnetic state ($T>T_{c}$), one would expect a linear $T$-dependence of 1/$\chi$ due to local moments.  Instead, we see a strong negative curvature.  Fitting the data in Fig.~\ref{susc1} in different temperature ranges between 200 and 350~K by the Curie-Weiss behavior in Eq.~(\ref{Curie}) gave the large value $\chi_0 = 3.8(2) \times 10^{-3}~{\rm cm^3/mol}$.  For a metal, $\chi_{0}$ has two contributions,  the diamagnetic conduction electron Landau orbital susceptibility $\chi_{\rm L}$ and the paramagnetic conduction electron Pauli spin susceptibility $\chi_{\rm P}$, in addition to the two contributions considered above for insulators, so that
\begin{equation}
\chi_0 = \chi_{\rm core} + \chi_{\rm VV} + \chi_{\rm L} + \chi_{\rm P}.
\label{Eqchi02}
\end{equation}
For transition metal compounds, the magnitude of $\chi_{\rm L}$ is small compared to $\chi_{\rm VV}$.  A typical value for $\chi_{\rm VV}$ is $\sim 1\times 10^{-4}$~cm$^3$ per mole of transition metal atoms, which gives $\chi_{\rm VV} \sim 3\times 10^{-4}$~cm$^3$ per mole of Sr$_{2}$Mn$_{2}$CuAs$_{2}$O$_2$.  Using the ionic model discussed above, we obtain $\chi_{\rm core} = -1.65
\times 10^{-4}$~cm$^{3}$/mol. We estimate $\chi_{\rm P}\sim 3 \times 10^{-4}$~cm$^3$/mol obtained
below from the heat capacity data analysis.  Adding these contributions gives $\chi_0 \sim 4 \times 10^{-4}$
cm$^{3}$/mol. The above fitted value is an order of magnitude larger than this estimated $\chi_{0}$ value and is hence unphysically large.  Therefore we next considered a model in which the observed strong positive curvature in $\chi^{-1}(T)$ in Fig.~\ref{susc1}(b) arises because the compound is a ferrimagnet.\cite{kittel1966}

From our neutron diffraction data below, we found that approximately equal amounts of Cu and Mn statistically occupy the metal atom positions in the $M$As layers, and that only Mn occupies the metal atom positions in the $M$O$_2$ layers.  One can therefore separate the magnetic atoms into A and B sublattices, where one of the sublattices is in the (Mn/Cu)As layer and the other sublattice is in the MnO$_2$ layer.  From the crystal structure in Fig.~\ref{structure}, one would expect three distinct interactions between the Mn and Mn/Cu atoms: an interaction within each sublattice and an interaction between sublattices, such as mentioned in the context of molecular field theory in Ref.~\onlinecite{kittel1966}.  However, having three interaction constants in addition to the unknown (average) Mn/Cu spin on one sublattice and Mn spin on the other sublattice as fitting parameters allows too many adjustable parameters in fitting the $\chi(T)$ data.  Therefore, as an \emph{effective} model, we set the intrasublattice (intralayer) interactions to zero and only consider bipartite interactions between magnetic atoms on the different A and B sublattices in adjacent layers.  The physical motivation for this choice is that that the adjacent Mn1 and Mn2/Cu2 layers will have distinctly different ordered moments/site and our model needs to differentiate these layers from each other.  

In the molecular field approximation, the dimensionless volume susceptibility $\chi_{\rm ferri}$ for such a ferrimagnet above its Curie temperature $T_{\rm C}$ is\cite{kittel1966}
\begin{equation}
\chi_{\rm ferri} = \frac{(C_{\rm A}+C_{\rm B})T-2T_{\rm C}^2/\mu}{T^{2}-T_{C}^{2}},
\label{chi0a}
\end{equation}
where the ferrimagnetic Curie temperature is
\begin{equation}
T_{\rm C} = \mu\sqrt{C_{\rm A}C_{\rm B}},
\label{TC}
\end{equation}
$\mu$ is a positive (antiferromagnetic) dimensionless molecular field coupling constant between the two sublattices and $C_{\rm A}$ and $C_{\rm B}$ are the Curie constants per unit volume for sublattices A and B, respectively.  The Curie constants and the saturation moment
$\mu_{\rm sat}$ per pair of A and B atoms are
\begin{equation}
C_{\rm A} = \frac{N_{\rm A}g^{2}\mu_{B}^{2} [S_{\rm A}(S_{\rm A}+1)]}{3k_{\rm B}(V_{\rm M}/2)},
\label{curie1}
\end{equation}
\begin{equation}
C_{\rm B} = \frac{N_{\rm A}g^{2}\mu_{B}^{2} [S_{\rm B}(S_{\rm B}+1)]}{3k_{\rm B}(V_{\rm M}/2)},
\label{curie2}
\end{equation}
\begin{equation}
\mu_{\rm sat} = g\mu_{\rm B}|S_{\rm A}-S_{\rm B}|,
\label{satmom}
\end{equation}
where $S_{\rm A}$ and $S_{\rm B}$ are the spins of the atoms on the A and B sublattices (or average spin in the case of the Mn/Cu layer), respectively, $g = 2$ is the $g$-factor which we take to be the same for both magnetic species A and B, and $V_{\rm M}/2$ is the volume per mole of A or B atoms which are assumed to be equal in number.

From our $M(H)$ measurements below, we obtain a reliable value for $\mu_{\rm sat}$, which allows us to eliminate one of the three variables $\mu$, $S_{\rm A}$ and $S_{\rm B}$ (we choose to eliminate $S_{\rm A}$ and we take $g = 2$) as a fitting parameter in Eq.~(\ref{chi0a}).  Since $\chi_{\rm ferri}$ in Eq.~(\ref{chi0a}) is the dimensionless volume susceptibility, our experimental molar $\chi(T)$ data in Fig.~\ref{susc1}(a) were converted to volume susceptibility by dividing by $V_{\rm M} = N_{\rm A}a^2c/4$ and were then plotted in Fig.~\ref{susc1}(b) and fitted by Eq.~(\ref{chi0a}) above 170~K\@.  We fixed $\mu_{\rm sat} = 2.20~\mu_{\rm B}$/f.u.\ as obtained from an $M(H)$ measurement at 1.8~K (see Fig.~\ref{MH1} below).  We first fitted the data by $\chi=\chi_{0}+\chi_{\rm ferri}$ and obtained the parameters $\chi_{0} = 1.06(7) \times 10^{-5}$ (which is equal to $0.99(7) \times 10^{-3}$~cm$^{3}$/mol), $S_{\rm B} = 0.26(1)$ and $\mu = 12600(300)$, which gives $T_{\rm C}=138(4)$~K\@. Using the value of $S_{\rm B}$ in Eq.~(\ref{satmom}), $S_{\rm A}$ was calculated to be 1.36(1). Thus the saturation moments per formula unit on layers B and A are
\bea
\mu_{\rm  sat} &=& 0.52~\mu_{\rm B}/{\rm B~layer} \nonumber\\
&&\hspace{1in}(\chi_0 \neq 0)\label{nonzerochi0}\\
\mu_{\rm  sat} &=& 2.72~\mu_{\rm B}/{\rm A~layer}. \nonumber 
\eea
In the ferrimagnetically-ordered state, these are antiparallel, giving a net saturation moment of $\mu = \mu_{\rm A~atom} - \mu_{\rm B~atom} = 2.20~\mu_{\rm B}$/f.u., by construction.  The $\chi_{0}$ value is somewhat larger than expected.  Therefore we carried out another fit by setting $\chi_{0} = 0$ and the resultant fitting parameters were $S_{\rm B} = 0.35(1)$, $\mu=9840(40)$, which gives $T_{\rm C} = 137(1)$~K\@.  From Eq.~(\ref{satmom}), $S_{\rm A}$ was calculated to be 1.45(1).  In this case, we have the saturation moments per formula unit
\bea
\mu_{\rm sat} &=& 0.70~\mu_{\rm B}/{\rm B~layer} \nonumber\\
&&\hspace{1in}(\chi_0 = 0)\label{zerochi0}\\
\mu_{\rm sat} &=& 2.90~\mu_{\rm B}/{\rm A~layer}. \nonumber 
\eea
Due to the highly simplified model, the saturation moments in Eqs.~(\ref{nonzerochi0}) and~(\ref{zerochi0}) should be considered to be semiquantitative only.  Indeed, from the results of the magnetic structure refinement by neutron diffraction in Table~\ref{22122Data} below, the Mn2 layer has $\mu_{\rm sat} = 2.2~\mu_{\rm B}$/f.u.\ and the Mn1 layer has $\mu_{\rm sat} = 3.9~\mu_{\rm B}$/f.u. 

We assume a Heisenberg exchange interaction between the sublattice A and B spins given by the Hamiltonian ${\cal H}= J\sum_{<ij>}\vec{S}_i \cdot \vec{S}_j$, where the sum is over nearest-neighbor A and B spin pairs.   We estimate the nearest-neighbor antiferromagnetic A-B exchange coupling $J$ from the molecular field coupling constant $\mu$ using\cite{kittel1966}
\begin{equation}
\mu = \frac{Jz\Omega}{g^{2}\mu_{\rm B}^{2}},
\label{exchange}
\end{equation}
where $\Omega=(V_{\rm M}/2)/N_{\rm A}$ is the volume per spin and $z = 8$ is the number of
nearest neighbor spins in the opposite sublattice in the two adjacent layers (see Fig.~\ref{structure}). Using the above values of $\mu$ yields $J/k_{\rm B} = 90(11)$~K\@.

\begin{figure}
\includegraphics [width=3.5in]{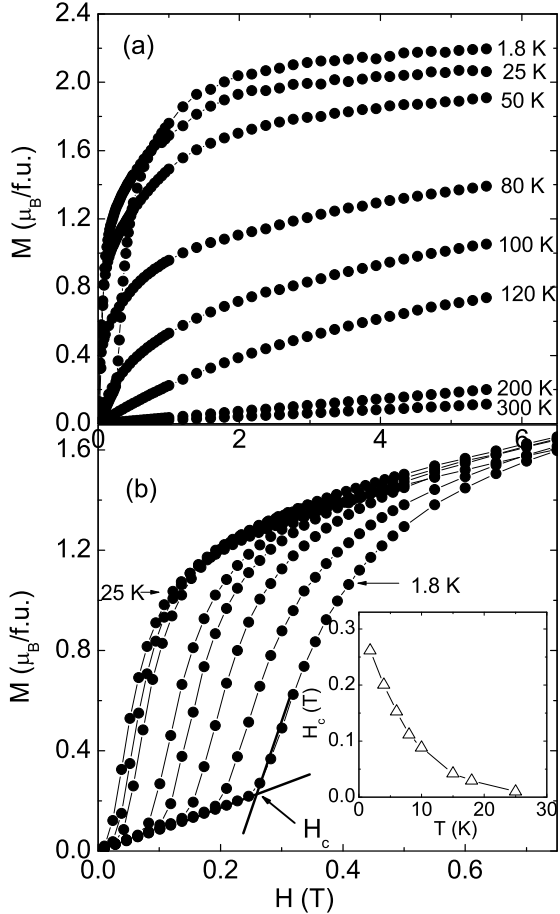}
\caption{\label{MH1} (a) Magnetization $M$ versus field $H$ isotherms
for Sr$_{2}$Mn$_{2}$CuAs$_{2}$O$_2$ at the indicated
temperatures. (b) Low temperature $M(H)$ at low $T$ and $H$.  The construction used to determine the metamagnetic critical field $H_{\rm c}$ is indicated.  Inset: $H_{\rm c}$ versus $T$.}
\end{figure}

\subsubsection*{Magnetization versus Applied Magnetic Field Measurements}

Figure~\ref{MH1} shows $M(H)$ isotherms
at different temperatures for Sr$_{2}$Mn$_{2}$CuAs$_{2}$O$_2$
measured up to $H = 5.5$~T\@.  At 300~K $M$ is proportional to $H$ but
below $150$~K $M(H)$ begins to show saturation.  At the lowest temperature
of $1.8$~K, $M$ almost saturates at $5$~T and the saturation moment
is $\mu_{\rm sat} \approx 2.2$~$\mu_{\rm B}$/f.u.\ if one assumes that $M$ saturates at high fields.  If one instead assumes a linear behavior at high fields and extrapolates the high field $M(H)$ to $H = 0$, one obtains a slightly lower value $\mu_{\rm sat} \approx 2.0$~$\mu_{\rm B}$/f.u.  These data suggest that Sr$_{2}$Mn$_{2}$CuAs$_{2}$O$_2$ is a ferromagnet or ferrimagnet in contrast to
Sr$_{2}$Zn$_{2}$MnAs$_{2}$O$_2$ which shows spin-glass ordering.  As
shown in Fig.~\ref{MH1}(b), $M(H)$ for
Sr$_{2}$Mn$_{2}$CuAs$_{2}$O$_2$ at $1.8$~K shows a pronounced step at
about $3$~kOe pointing towards a spin-flop type metamagnetic
transition at a critical field $H_{\rm c}$. As shown in the inset of
Fig.~\ref{MH1}(b), $H_{\rm c}$ decreases with increasing $T$ and is
completely suppressed at 25~K, which is far below the Curie
temperature $T_{\rm C}$. Since the material is a ferrimagnet $H_{\rm
c}$ may therefore be associated with domain wall depinning.

\begin{figure}
\includegraphics [width=3in]{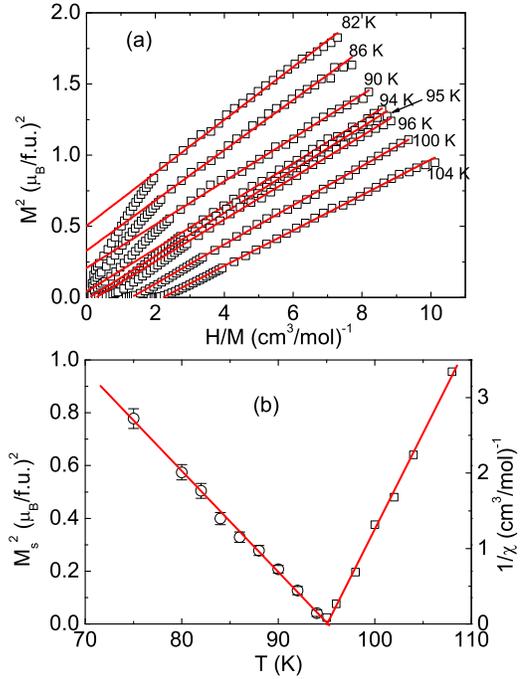}
\caption{\label{Arrott} (Color online) (a) Arrott plots of the square of the magnetization $M^{2}$ versus the ratio of the applied magnetic field to the magnetization $H/M$ for Sr$_{2}$Mn$_{2}$CuAs$_{2}$O$_2$ at representative temperatures $T$ near $T_{\rm C}$. The solid red lines are linear fits to the data for $H \geq 1$~T and are extrapolated to $H/M = 0$. (b) Square of the spontaneous magnetization $M_{s}^{2}$ (left-hand scale) and the inverse of the intrinsic zero-field magnetic susceptibility $1/\chi$ (right-hand axis) versus $T$, derived from the fits in (a). The solid red lines are linear fits to the respective data.}
\end{figure}

In order to
further study the magnetic behavior, we measured $M(H)$
isotherms in the vicinity of $T_{\rm C}$.  With the help of Arrott
plots\cite{arrott1957} in which the square of the magnetization $M^{2}$
in an applied field $H$ is plotted as a function of $H/M$ for fixed
temperatures $T$, we estimated $T_{\rm C}$. This is a very useful and standard method for establishing
the presence of ferromagnetic/ferrimagnetic order and also for an accurate
determination of $T_{\rm C}$. This approach is based on the Weiss
molecular-field theory which predicts that such plots yield straight
line isotherms with an intercept which tends to zero as $T$
approaches $T_{\rm C}$ from above. However experimentally such Arrott plots can exhibit considerable curvature arising from non-mean-field behavior. Therefore generalized Arrott plots have been used where $M^{1/\beta}$ is plotted against
$(H/M)^{1/\gamma}$.\cite{knafo2009, hadimani2008, castano1999} Based
on the values of the critical exponents $\beta$ and $\gamma$ values that give
linear plots, the universality class of the magnetic system can be inferred.

Figure~\ref{Arrott}(a) shows Arrott plots of $M^{2}$ vs $H/M$ near
$T_{\rm C}$ for Sr$_{2}$Mn$_{2}$CuAs$_{2}$O$_{2}$. The data are well
fitted by linear behaviors for fields higher than $1$~T\@. Data at low fields are ignored in such plots because those data can be unduly affected by a small fraction of the sample.  For $T > T_{\rm C}$ one has 
\begin{equation}
M^2(T) = A(T)\left[\frac{H}{M(H,T)} - \frac{1}{\chi(T)}\right],\ \ \ \ (T > T_{\rm C})
\label{EqArrott1}
\end{equation}
from which one can infer the inverse of the intrinsic zero-field susceptibility $\chi^{-1}(T)$ as the intercept on the horizontal $H/M$ axis as shown in Fig.~\ref{Arrott}(b) (right-hand axis).  A linear extrapolation of the $1/\chi$ data to $1/\chi = 0$ gives $T_{\rm C} = 95.1(2)$~K\@.  Fitting the $\chi^{-1}(T)$ data in Fig.~\ref{Arrott}(b) by a Curie-Weiss law gives the Curie constant and effective moment as
\bea
C &=& 3.86(9)~{\rm cm^3~K/mol}\nonumber\\
\label{mueffC}\\
\mu_{\rm eff} &=&  5.55(6)~\mu_{\rm B}/{\rm f.u.}\nonumber
\eea
Anticipating the results of the magnetic structure refinement by neutron diffraction in Table~\ref{22122Data} below,  we have one Mn1 spin $S = 2$ and two Mn2 \emph{site} spins $S = 1/2$ per formula unit, giving an effective moment of 5.48~$\mu_{\rm B}$/f.u.\ assuming $g = 2$, in excellent agreement with the value in Eq.~(\ref{mueffC}).

For $T < T_{\rm C}$, one has 
\begin{equation}
M^2(T) = M_{\rm s}^2(T) + A\frac{H}{M},\ \ \ \ (T < T_{\rm C}),
\label{EqArrott2}
\end{equation}
from which one can infer the square of the spontaneous magnetization $M^2_{\rm s}(T)$ from the intersection of the fitted line with the vertical $M^2$ axis as shown in Fig.~\ref{Arrott}(b) (left-hand axis).  The susceptibility at $T > T_{\rm C}$ and saturation magnetization at $T < T_{\rm C}$ of a ferrimagnet or ferromagnet are expected to follow critical behaviors $1/\chi = (T/T_{\rm C} - 1)^\gamma$ and $M_{s} \sim (1-T/T_{\rm C})^\beta$, respectively.\cite{castano1999} As shown in
Fig.~\ref{Arrott}(b), $M_{s}^{2}$ versus $T$ is linear 
indicating that the critical exponent $\beta \approx 0.50$ and 
the linear $T$-dependence of $1/\chi$ gives $\gamma = 1$ which
are both consistent with the predictions of mean-field theory.\cite{knafo2009}  From these results 
Sr$_{2}$Mn$_{2}$CuAs$_{2}$O$_2$ appears to be a good example of a
mean-field ferrimagnet.  Therefore, it is very surprising that there is no discernable heat capacity anomaly associated with the transition at $T_{\rm C} = 95$~K [see Fig.~\ref{Arrott}(b)], as shown in the next section.

\subsection{Heat Capacity Measurements}

\begin{figure}
\includegraphics [width=3in]{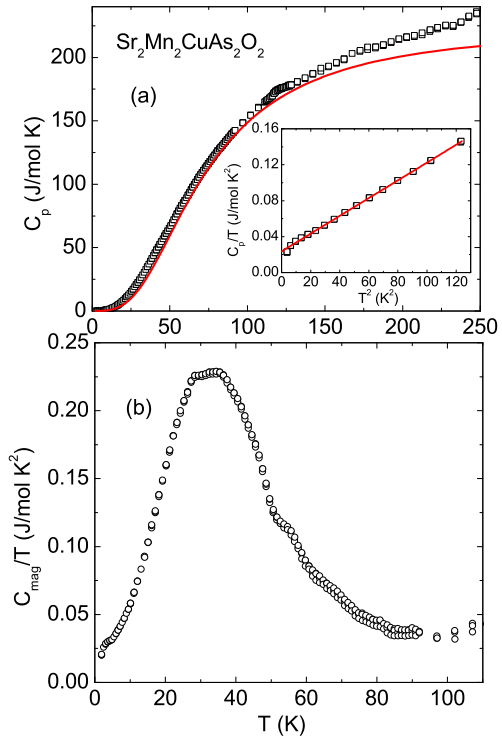}
\caption{\label{cp1} (Color online) (a) Heat capacity
$C_{\rm p}$ versus temperature $T$ for Sr$_{2}$Mn$_{2}$CuAs$_{2}$O$_2$. The
solid red curve is a fit of the Debye function to the data around 100~K which is taken as an approximation to the lattice heat capacity.  Inset: $C_{\rm p}/T$ vs. $T^{2}$ below 11~K\@.  The red line is a linear fit to the data from 1.8 to 11~K\@. (b) Magnetic contribution to the heat capacity divided by temperature, $C_{\rm mag}/T$, versus $T$.}
\end{figure}

\begin{table}
\caption{\label{CpData} Linear specific heat coefficients $A$ and Debye temperatures $\Theta_{\rm D}$ for Sr$_{2}$Mn$_{3}$As$_{2}$O$_{2}$, Sr$_{2}$Zn$_{2}$MnAs$_{2}$O$_{2}$ and Sr$_{2}$Mn$_{3}$As$_{2}$O$_{2}$ obtained from fitting the low-temperature specific heat data by Eq.~(\ref{CpFitEq}).  The large $A$ value for Sr$_{2}$Zn$_{2}$MnAs$_{2}$O$_{2}$ may not be the low-$T$ limit because $C(T)$ for this sample showed a sharp downturn below 4~K\@.}
\begin{ruledtabular}
\begin{tabular}{l|cc}
Compound & $A$ & $\Theta_{\rm D}$ \\
& (mJ/mol~K$^2$) & (K)\\
\hline
Sr$_{2}$Mn$_{2}$CuAs$_{2}$O$_{2}$ & 23.6(3) & 260(1)\\
Sr$_{2}$Zn$_{2}$MnAs$_{2}$O$_{2}$ & 73.0(4) & 252(1)\\
Sr$_{2}$Mn$_{3}$As$_{2}$O$_{2}$ & 13.1(3) & 263(1)\\
\end{tabular}
\end{ruledtabular}
\end{table}

The heat capacity at constant pressure $C_{\rm p}$ at $H = 0$ is plotted versus $T$ in
Figs.~\ref{cp1}(a), \ref{cp2}(a), and \ref{cp3} for 
Sr$_{2}$Mn$_{2}$CuAs$_{2}$O$_2$, Sr$_{2}$Zn$_{2}$MnAs$_{2}$O$_2$, and
Sr$_{2}$Mn$_{3}$As$_{2}$O$_2$, respectively. The values of
$C_{\rm p}$ at room 250~K are about 240 and 225~J/mol~K for Sr$_{2}$Mn$_{2}$CuAs$_{2}$O$_2$ and 
Sr$_{2}$Zn$_{2}$MnAs$_{2}$O$_2$, 
respectively. These values are 
comparable with the Dulong-Petit classical lattice heat capacity at constant volume $C_{\rm V} = 27R =
225$~J/mol~K.\cite{kittel1966} The observed value that is somewhat larger than the Dulong-Petit value may be due to the fact that the compound is measured at constant pressure rather than constant volume and/or to a significant magnetic contribution being present in addition to the lattice contribution.  

\subsubsection{\rm Sr$_{2}$Mn$_{2}$CuAs$_{2}$O$_2$}

As shown in the inset of Fig.~\ref{cp1}(a) $C_{\rm p}(T)/T$ versus $T^{2}$ for Sr$_{2}$Mn$_{2}$CuAs$_{2}$O$_2$ is almost linear at low temperatures (2~K $\leq T \leq $ 11~K).  The resistivity data for this compound in Fig.~\ref{resist1} suggested a metallic ground state, so we fitted the data at low temperatures by the expression 
\be 
\frac{C_{\rm p}(T)}{T} = A + \beta T^2,
\label{CpFitEq}
\ee
where the first term is interpreted for this sample as the Sommerfeld electronic specific heat coefficient due to conduction electrons and the second term is the low-temperature limit of the lattice heat capacity.  The resultant $A$ value is listed in Table~\ref{CpData}. 

From the value of $\beta$ one can estimate the Debye temperature $\Theta_{\rm D}$ using the
expression\cite{kittel1966}
\begin{equation}
\Theta_{\rm D}=\left( \frac{12\pi ^{4}Rn}{5\beta}\right)^{1/3}  ,
\label{theta}
\end{equation}
where $R$ is the molar gas constant and $n$ is the number of atoms
per formula unit ($n=9$ for our compounds). The $\beta$ value
yields the value of $\Theta_{\rm D}$ listed in Table~\ref{CpData}.

The density of states at the Fermi energy for both spin directions
$N(E_{\rm F})$ can be estimated from the value of $A$ using the
relation\cite{kittel1966}
\begin{equation}
A =\frac{\pi ^{2}}{3}%
k_{\rm B}^{2}N(E_{\rm F})\left( 1+\lambda _{\rm ep}\right)
\label{gama}
\end{equation}
where $\lambda _{\rm ep}$ is the electron-phonon coupling constant.
As a first approximation we set $\lambda _{\rm ep}=0$, which gives
$N(E_{\rm F})\simeq 10.0$~states/(eV f.u.).  From $N(E_{\rm F})$ we calculated
the Pauli spin susceptibility $\chi_{\rm P}$ using\cite{kittel1966}
\begin{equation}
\chi_{\rm P}=\mu_{\rm B}^{2} N(E_{\rm F})
\end{equation}
where $\mu_{\rm B}$ is the Bohr magneton. This gives $\chi_{\rm P}\simeq
3.24 \times 10^{-4}$~cm$^3$/mol.

\subsubsection{\rm Sr$_{2}$Zn$_{2}$MnAs$_{2}$O$_2$ and Sr$_{2}$Mn$_{3}$As$_{2}$O$_2$}

\begin{figure}
\includegraphics [width=3in]{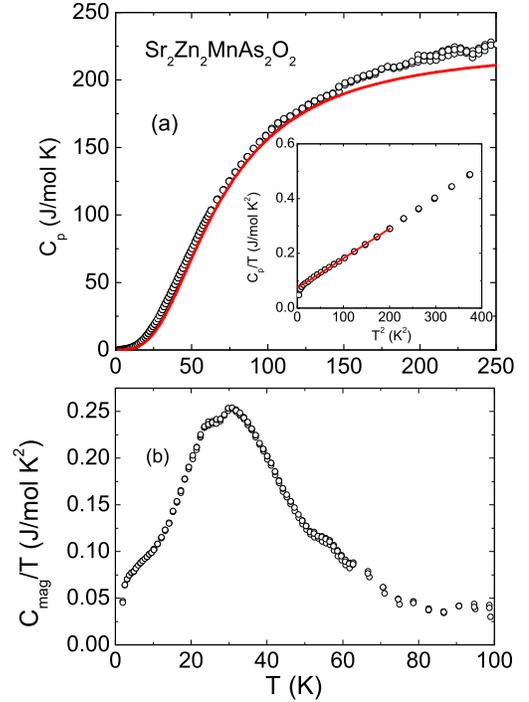}
\caption{\label{cp2} (Color online) (a) Heat capacity
$C_{\rm p}$ versus temperature $T$ for Sr$_{2}$Zn$_{2}$MnAs$_{2}$O$_2$. The
solid red curve is a fit of the Debye function to the data around 100~K which is taken as an approximation to the lattice heat capacity.  Inset: $C_{\rm p}/T$ vs. $T^{2}$ below 20~K\@.  The red line is a linear fit to the data from 4 to 14~K\@. The data below $4$~K show a sharp decrease.  The data start to deviate from linear behavior above $14$~K\@.  (b) Magnetic contribution to the heat capacity divided by temperature, $C_{\rm mag}/T$, versus $T$.}
\end{figure}

\begin{figure}
\includegraphics [width=3in]{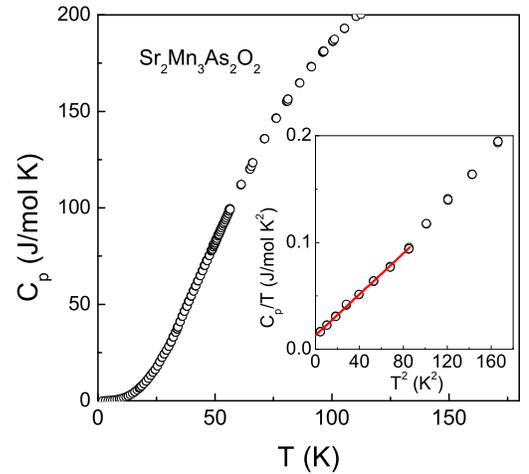}
\caption{\label{cp3} (Color online) Heat capacity $C_{\rm p}$ vs.
temperature $T$ for Sr$_{2}$Mn$_{3}$As$_{2}$O$_2$. Inset: $C_{\rm p}/T$ vs. $T^{2}$.  The solid red line is a linear fit from 1.8 to~9~K\@.}
\end{figure}

For Sr$_{2}$Zn$_{2}$MnAs$_{2}$O$_2$, $C_{\rm p}(T)/T$ versus $T^{2}$ shows a sharp decrease
below about 4~K but a linear region was observed just above this
temperature (4~K $\leq T \leq 14$~K) [inset of
Fig.~\ref{cp2}(a)]. For Sr$_{2}$Mn$_{3}$As$_{2}$O$_2$ which is also an insulator, $C_{\rm p}(T)/T$ versus $T^{2}$ is almost linear from our low temperature limit of 2~K up to 9~K (inset of Fig.~\ref{cp3}).  To parametrize the data, we fitted the $C_{\rm p}(T)/T$ vs. $T^{2}$ data in these linear regimes by Eq.~(\ref{CpFitEq}).  The parameters obtained from the fits are shown in Table~\ref{CpData}.  For these compounds, since the $\rho(T)$ measurements indicated insulating ground states, the $A$ values may be related to the spin-glass like behavior observed in $\chi(T)$.\cite{matsushita2000}   In Sr$_{2}$Mn$_{3}$As$_{2}$O$_2$, since the MnAs sublattice orders 
at 340~K, in the ordered state the magnetic part of the heat capacity 
may contain a $T^{3}$ term due to spin-wave excitations, and our 
experimental $\beta$ would then have a contribution from the spin-wave 
excitations. Thus our estimated $\Theta_{\rm D}$ gives a 
lower limit of the Debye temperature in Sr$_{2}$Mn$_{3}$As$_{2}$O$_2$. These 
estimated $\Theta_{\rm D}$ values are much higher than that reported for 
Ba$_{2}$Zn$_{2}$MnAs$_{2}$O$_2$ ($\Theta_{\rm D}= 111$~K)\cite{matsushita2000} 
but are comparable to those for BaMn$_{2}$As$_{2}$
(Refs.~\onlinecite{an2009} and~\onlinecite{singh2009}) and
(Ba,Sr)Ru$_{2}$As$_{2}$.\cite{nath2009}

\subsubsection{Magnetic Heat Capacity Contributions}

In an attempt to extract the magnetic heat capacity $C_{\rm mag}(T)$ from the observed data, we approximate the lattice contribution $C_{\rm latt}(T)$ by the Debye function\cite{kittel1966}
\begin{equation}
C_{\rm latt}(T) = 9R\left( \frac{T}{\Theta_{\rm D}}
\right)^{3}\int_{0}^{\Theta_{\rm D}/T}\frac{x^{4}e^{x}}{\left(
e^{x}-1\right)}dx.
\label{Debye}
\end{equation}
The solid curves in Figs.~\ref{cp1}(a) and~\ref{cp2}(a) are plots of $C_{\rm
latt}(T)$ obtained by fitting the high temperature data, yielding $\Theta_{\rm D} = 300$~K and 280~K for Sr$_{2}$Mn$_{2}$CuAs$_{2}$O$_2$ and Sr$_{2}$Zn$_{2}$MnAs$_{2}$O$_2$,
respectively.   These $\Theta_{\rm D}$ values are somewhat larger than the values in Table~\ref{CpData} estimated from our \emph{low}-temperature heat capacity measurements.  The magnetic contribution $C_{\rm mag}(T)$ is obtained by subtracting the respective calculated lattice contribution from the observed data.  The resulting  $C_{\rm mag}(T)/T$ is plotted versus $T$ in Figs.~\ref{cp1}(b) and~\ref{cp2}(b), respectively.  We did not observe
any clear anomaly in $C_{\rm mag}(T)$ associated with any magnetic
transition as inferred from the above $\chi(T)$ measurements. The magnetic entropies $S_{\rm mag}(T) = \int_{1.8\,{\rm K}}^T  [C_{\rm mag}(T^\prime)/T^\prime]dT^\prime$ at
$T =100$~K are 5.05 and 11~J/(mol~Mn~K) for
Sr$_{2}$Mn$_{2}$CuAs$_{2}$O$_2$ and Sr$_{2}$Zn$_{2}$MnAs$_{2}$O$_2$,
respectively.

We were not able to fit the high-$T$ $C_{\rm p}(T)$ data for Sr$_{2}$Mn$_{3}$As$_{2}$O$_2$ adequately by Eq.~(\ref{Debye}).  This may be due to the magnetic ordering of the MnAs sublattice at 340~K which could have a significant magnetic contribution thereby enhancing the total $C_{\rm p}$ at high temperatures.

From the neutron diffaction solution of the combined crystallographic and magnetic structures at 4~K for Sr$_{2}$Mn$_{2}$CuAs$_{2}$O$_2$ in Table~\ref{22122Data} below, the $n_1 = 1$ Mn1 atoms/f.u.\ have ordered moments of 3.89(11)~$\mu_{\rm B}$/Mn corresponding to spin $S_1 = 2$ with $g = 2$, and each of the $n_2 = 2$ Mn2 \emph{sites}/f.u.\ has a magnetic moment of 1.10(7)~$\mu_{\rm B}$/Mn corresponding to spin $S_2 = 1/2$.  Therefore the entropy of the disordered spins at high temperatures is calculated to be
\be
S_{\rm calc} = \sum_{i = 1}^2 n_iR\ln(2S_i+1) = 3.00R = 24.9~{\rm \frac{J}{mol~K}},
\label{EntropyCalc}
\ee
where $R$ is the molar gas constant.  This is much larger than the value of 10.1~J/mol~K at 100~K estimated above for the magnetic component of the heat capacity.  Similarly, for Sr$_{2}$Zn$_{2}$MnAs$_{2}$O$_2$, if we assume a spin between 2 and 5/2 for the Mn, the predicted disordered entropy is between 13.4 and 14.9~J/mol~K, which is again somewhat larger than the value $S_{\rm mag} = 11$~J/(mol~K) inferred above.  These discrepancies likely result from inaccurate estimates of the respective lattice heat capacities that were used to obtain the magnetic heat capacity from the observed values.  It would be useful to synthesize and measure the heat capacity of a nonmagnetic reference compound to obtain a better estimate of the lattice heat capacity versus temperature.

\subsection{Neutron Powder Diffraction Measurements}

Neutron powder diffraction measurements were undertaken to answer
two key questions regarding the Sr$_{2}$Mn$_{2}$CuAs$_{2}$O$_{2}$
compound. First, it is important to identify the site(s)
in the lattice occupied by the Cu. The large difference between the coherent scattering lengths\cite{scattLengths} $b_{\rm coh}$
for Mn ($b_{\rm coh} = -3.75$~fm) and Cu ($b_{\rm coh} = 7.72$~fm) provides
excellent contrast for this purpose.  Second, as described above,
magnetization measurements suggest that
Sr$_{2}$Mn$_{2}$CuAs$_{2}$O$_{2}$ is ferrimagnetic, in contrast to
the G-type antiferromagnetic ordering found for Mn2 in Sr$_{2}$Mn$_{3}$As$_{2}$O$_{2}$, where the two members of all nearest neighbor spin pairs are antiparallel.
In addition, we carried out measurements on a powder sample of Sr$_{2}$Mn$_{3}$As$_{2}$O$_{2}$ for comparison with previous work and with our data for Sr$_{2}$Mn$_{2}$CuAs$_{2}$O$_{2}$.  The microscopic details of the magnetic structure of
Sr$_{2}$Mn$_{2}$CuAs$_{2}$O$_{2}$, and how it differs from the
Sr$_{2}$Mn$_{3}$As$_{2}$O$_{2}$ parent compound are, again, best
probed by neutron diffraction. 

\subsubsection{{\rm Sr}$_{2}${\rm Mn}$_{3}${\rm As}$_{2}${\rm O}$_{2}$}

\begin{figure}[tbp]
\includegraphics[width=3.3in]{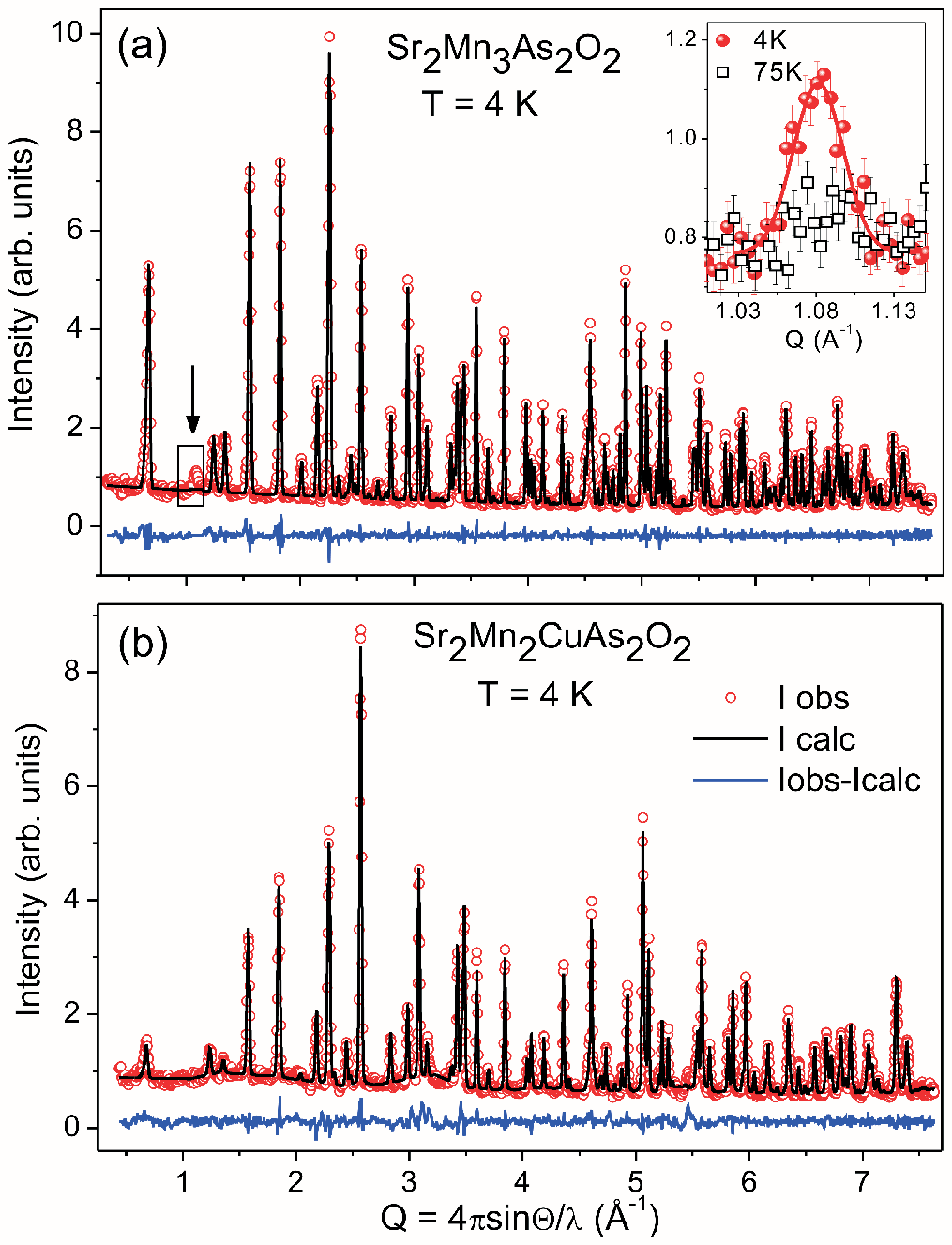}
\caption{\label{rietveld}(Color online) Neutron diffraction data collected at temperature 
$T = 4$~K (open circles) and calculated profiles (solid line) for 
(a) Sr$_{2}$Mn$_{3}$As$_{2}$O$_{2}$ and (b) Sr$_{2}$Mn$_{2}$CuAs$_{2}$O$_{2}$. 
The difference between observed and calculated profiles is shown at the bottom 
of each panel. The inset in (a) displays the (100)$_{\rm mag}$ magnetic peak which appears below 75~K for Sr$_{2}$Mn$_{3}$As$_{2}$O$_{2}$. The position of this peak is indicated by the vertical arrow located towards the left in the main figure.}
\end{figure}

\begin{table*}
\caption{\label{2322Data} Crystallographic and magnetic neutron powder diffraction refinement data for Sr$_{2}$Mn$_{3}$As$_{2}$O$_{2}$, body-centered-tetragonal, space group \emph{I}4/\emph{mmm} (No.~139), $Z = 2$ formula units/unit cell.  The Mn1 atoms are in MnO$_2$ layers and the Mn2 atoms are in  Mn$_2$As$_2$ layers.}
\begin{ruledtabular}
\begin{tabular}{lcccc}
$T$
& 375~K
& 300~K
& 150~K
& 4~K \\
\hline
$a$ (\AA)
& 4.14453(2)
& 4.14079(4)
& 4.13438(3)
& 4.13104(4)
\\
$c$ (\AA)
& 18.8530(1)
& 18.8217(2)
& 18.7672(1)
& 18.7382(2) \\
Unit cell volume (\AA$^3$)
& 323.841(3)
& 322.721(5)
& 320.789(4)
& 319.776(6)\\
\hline
Sr in 4\emph{e} (0 0 $z$)  \\
$z$
& 0.41379(6)
& 0.41362(6)
& 0.41358(6)
& 0.41358(7) \\
$B_{\rm iso}$
& 0.81(2)
& 0.68(6)
& 0.34(3)
& 0.26(3) \\
\hline
Mn1 in 2\emph{a} (0 0 0) \\
$B_{\rm iso}$
& 0.82(6)
& 0.78(4)
& 0.47(5)
& 0.33(7) \\
Ordered moment ($\mu_{\rm B}$/Mn1)
& ---
& ---
& 0
& ? \\
\hline
Mn2 in 4\emph{d} (0 1/2 1/4) \\
$B_{\rm iso}$
& 0.99(4)
& 0.40(5)
& 0.41(4)
& 0.36(4) \\
Ordered moment ($\mu_{\rm B}$/Mn2) & --- & 2.10(3)
& 3.18(3)
& 3.50(4) \\
\hline
As in 4\emph{e} (0 0 $z$)\\
$z$
& 0.16900(7)
& 0.16892(7)
& 0.16893(6)
& 0.16894(8) \\
$B_{\rm iso}$
& 0.80(2)
& 0.61(3)
& 0.30(3)
& 0.26(3) \\
\hline
O in 4\emph{c} (0 1/2 0)\\
$B_{\rm iso}$
& 1.01(3)
& 0.89(3)
& 0.55(3)
& 0.43(3) \\
\hline
$R_{\rm Bragg}$ (\%)
& 5.35
& 4.87
& 4.37
& 4.78 \\
$R_{\rm Magnetic}$ (\%)
& --- & 4.38
& 2.77
& 3.1 \\
$\chi^2$
& 1.71
& 1.89
& 2.54
& 5.41 \\

\end{tabular}
\end{ruledtabular}
\end{table*}

In Fig.~\ref{rietveld}(a) we plot the neutron diffraction data at temperature $T = 4$~K, along with the fit that
results from the refinement that includes the crystallographic and
G-type magnetic structure as well as a small amount (2--3\%) of MnO
impurity phase. The crystallographic parameters from our refinements
of Sr$_{2}$Mn$_{3}$As$_{2}$O$_{2}$ are listed in Table~\ref{2322Data}. The error
estimates in this table represent only the statistical errors
associated with the refinement itself.  Consistent with previous
work, we found no significant changes in crystal structure with
temperature.

\begin{figure}[]
\includegraphics[width=3.1in]{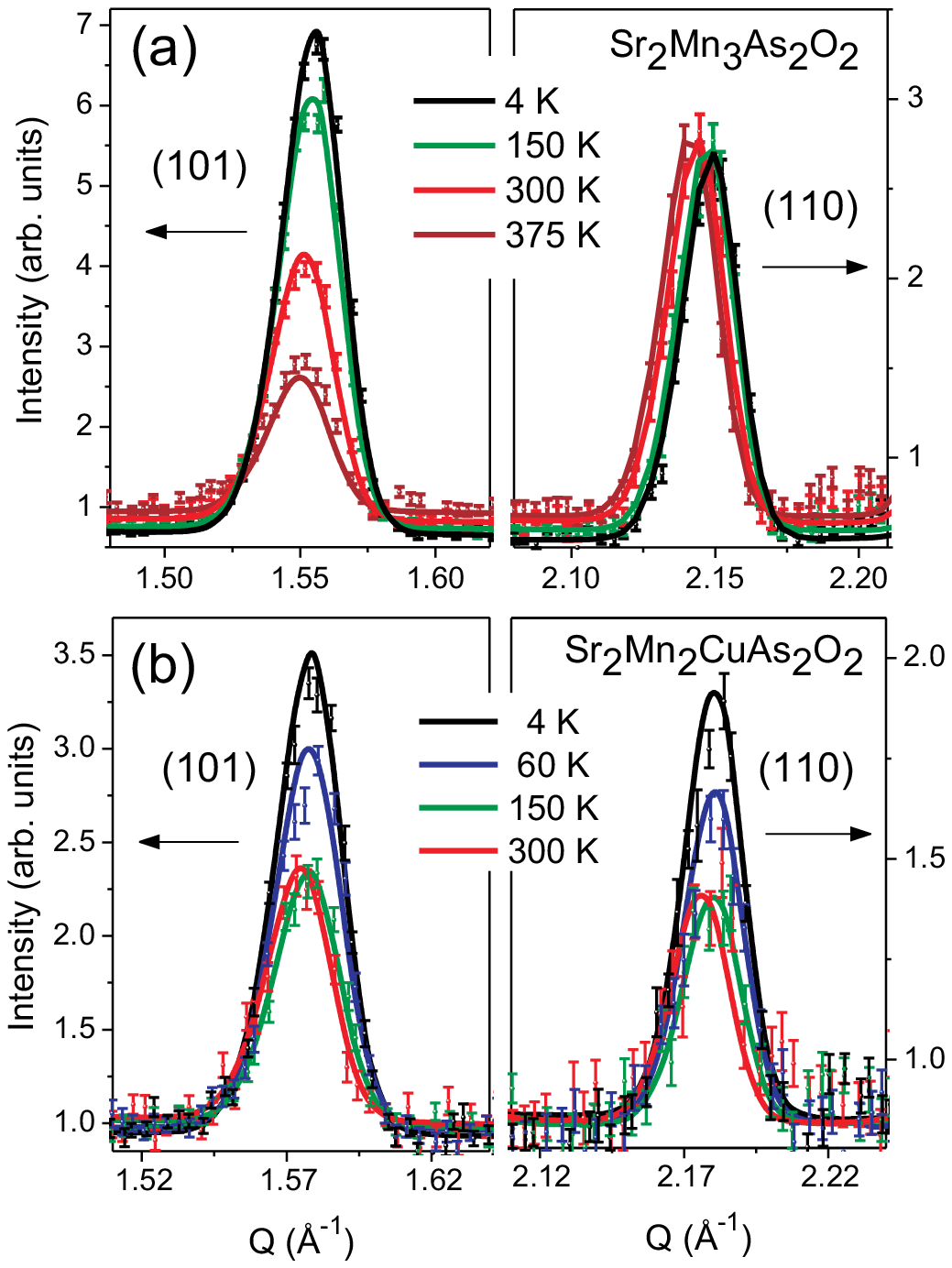}
\caption{\label{bragg}(Color online) Temperature dependence of the intensity of the (101) and (110) 
diffraction peaks for (a) Sr$_{2}$Mn$_{3}$As$_{2}$O$_{2}$ and (b) Sr$_{2}$Mn$_{2}$CuAs$_{2}$O$_{2}$}.
\end{figure}

Focusing now on the magnetic structure, in Fig.~\ref{bragg}(a) we show that
as temperature is decreased below 300~K the intensity of the (101)
diffraction peak increases, while the intensity of the (110) peak
remains constant, again consistent with the G-type antiferromagnetic
ordering of the Mn2 site as described previously,\cite{brock1996} and
illustrated in Fig.~\ref{magstr}(a). The ordered moment at our base
temperature of 4~K is 3.50(4)~$\mu_{\rm B}$ per Mn2 ion.  

As shown in the inset of Fig.~\ref{rietveld}(a), at 4~K we also
find evidence of a change in the magnetic structure with the
appearance of a new peak, identified by Brock et al.\ as the (100)
reflection associated with the magnetic unit cell described by
$a_{\rm mag} = \sqrt{2}a$ and $c_{\rm mag} = c$ for
Sr$_{2}$Mn$_{3}$As$_{2}$O$_{2}$. The new magnetic peak was previously assigned to a 
quasi-2D-ordering of the Mn1 sublattice.\cite{brock1996}  However, we observed no additional reflections
at the expected higher index positions [e.g.\ (110)$_{\rm mag}$ and (112)$_{\rm mag}$]. 
Since no model was found to fully explain the observed scattering data, the above refinement of the magnetic structure at 4~K 
was conducted by excluding the data points associated with the new peak.  Single crystal 
neutron measurements would be useful to resolve the nature of this additional reflection and
its relationship to the magnetic structure at low temperature.

\subsubsection{{\rm Sr}$_{2}${\rm Mn}$_{2}${\rm Cu}{\rm As}$_{2}${\rm O}$_{2}$}

\begin{table*}
\caption{\label{22122Data} Crystallographic and magnetic neutron powder diffraction refinement data for Sr$_{2}$Mn$_{2}$CuAs$_{2}$O$_{2}$, body-centered-tetragonal, space group \emph{I}4/\emph{mmm} (No.~139), $Z = 2$ formula units/unit cell.  The Mn1 atoms are in MnO$_2$ layers and the Mn2/Cu atoms are in (Mn/Cu)$_2$As$_2$ layers.}
\begin{ruledtabular}
\begin{tabular}{lcccc}
$T$
& 300~K
& 120~K
& 60~K
& 4~K \\
\hline
$a$ (\AA)
& 4.07913(6)
& 4.07198(5)
& 4.07086(5)
& 4.07106(5)\\
$c$ (\AA)
& 18.5826(3)
& 18.5216(2)
& 18.4861(2)
& 18.4784(2)\\
Unit cell volume (\AA$^3$)
& 309.201(8)
& 307.108(7)
& 306.351(7)
& 306.253(7)\\
\hline
Sr in 4\emph{e} (0 0 $z$)  \\
$z$
& 0.4094(1)
& 0.40933(9)
& 0.40932(9)
& 0.40913(9)\\
$B_{\rm iso}$
& 0.57(4)
& 0.32(4)
& 0.16(3)
& 0.10(3)\\
\hline
Mn1 in 2\emph{a} (0 0 0) \\
$B_{\rm iso}$
& 0.62(9)
& 0.37(8)
& 0.16(8)
& 0.13(8)\\
Ordered moment ($\mu_{\rm B}$/Mn1)
& ---
& ---
& 2.98(13)
& 3.89(11)\\
\hline
Mn2/Cu in 4\emph{d} (0 1/2 1/4) \\
Cu occupancy 
& 37(1)\%
& 37\%
& 37\%
& 37\% \\
$B_{\rm iso}$
& 0.65(5)
& 0.40(5)
& 0.22(5)
& 0.16(5)\\
Ordered moment ($\mu_{\rm B}$/Mn2 site)\footnotemark[1] & --- & --- 
& 0.91(10)& 1.10(7)\\
Ordered moment ($\mu_{\rm B}$/Mn2 atom)\footnotemark[2] & --- & --- 
& 1.43(17) & 1.72(13)\\
\hline
As in 4\emph{e} (0 0 $z$)\\
$z$
& 0.1684(1)
& 0.1683(1)
& 0.1684(1)
& 0.1684(1) \\
$B_{\rm iso}$
& 0.58(4)
& 0.35(4)
& 0.23(4)
& 0.26(4)\\
\hline
O in 4\emph{c} (0 1/2 0)\\
$B_{\rm iso}$
& 0.91(5)
& 0.64(5)
& 0.53(5)
& 0.47(5) \\
\hline
$R_{\rm Bragg}$ (\%)
& 6.04
& 5.23
& 5.81
& 5.38 \\
$R_{\rm Magnetic}$ (\%)
& --- & --- 
& 12.4
& 10.3\\
$\chi^2$
& 2.37
& 2.80
& 2.65
& 3.28 \\

\end{tabular}
\end{ruledtabular}
\footnotetext[1]{Measured.}
\footnotetext[2]{Inferred from the 63\% occupation of the Mn2 sites by Mn and 37\% by nonmagnetic Cu.}
\end{table*}

Moving now to the Sr$_{2}$Mn$_{2}$CuAs$_{2}$O$_{2}$ sample, Fig.~\ref{rietveld}(b) and Table~\ref{22122Data} describe the results of our refinements of the
crystallographic and magnetic structures of this compound. Similar to what was found
for Sr$_{2}$Mn$_{3}$As$_{2}$O$_{2}$, the diffraction pattern
indicates the presence of MnO but, in addition, a small amount of a
second phase that could not be identified also appears.
Nevertheless, a reasonable refinement of the crystallographic and
magnetic phase parameters was obtained. As Table~\ref{22122Data} shows, the most
striking result of the crystallographic refinement is that the Cu is
found only in the Mn2 site, associated with the ${\rm Mn_2As_2}$ layers in the
structure.  Attempts to substitute Cu in the Mn1 sites resulted in
significantly poorer refinement $R$-factors.  We further note that,
given the stoichiometry of the sample, the Cu occupancy on the Mn2
site is somewhat lower than the nominal value (37\%
vs.\ 50\%), but this decrease may be associated with the
presence of the second impurity phase in this sample.

\begin{figure}[]
\includegraphics[width=3.5in]{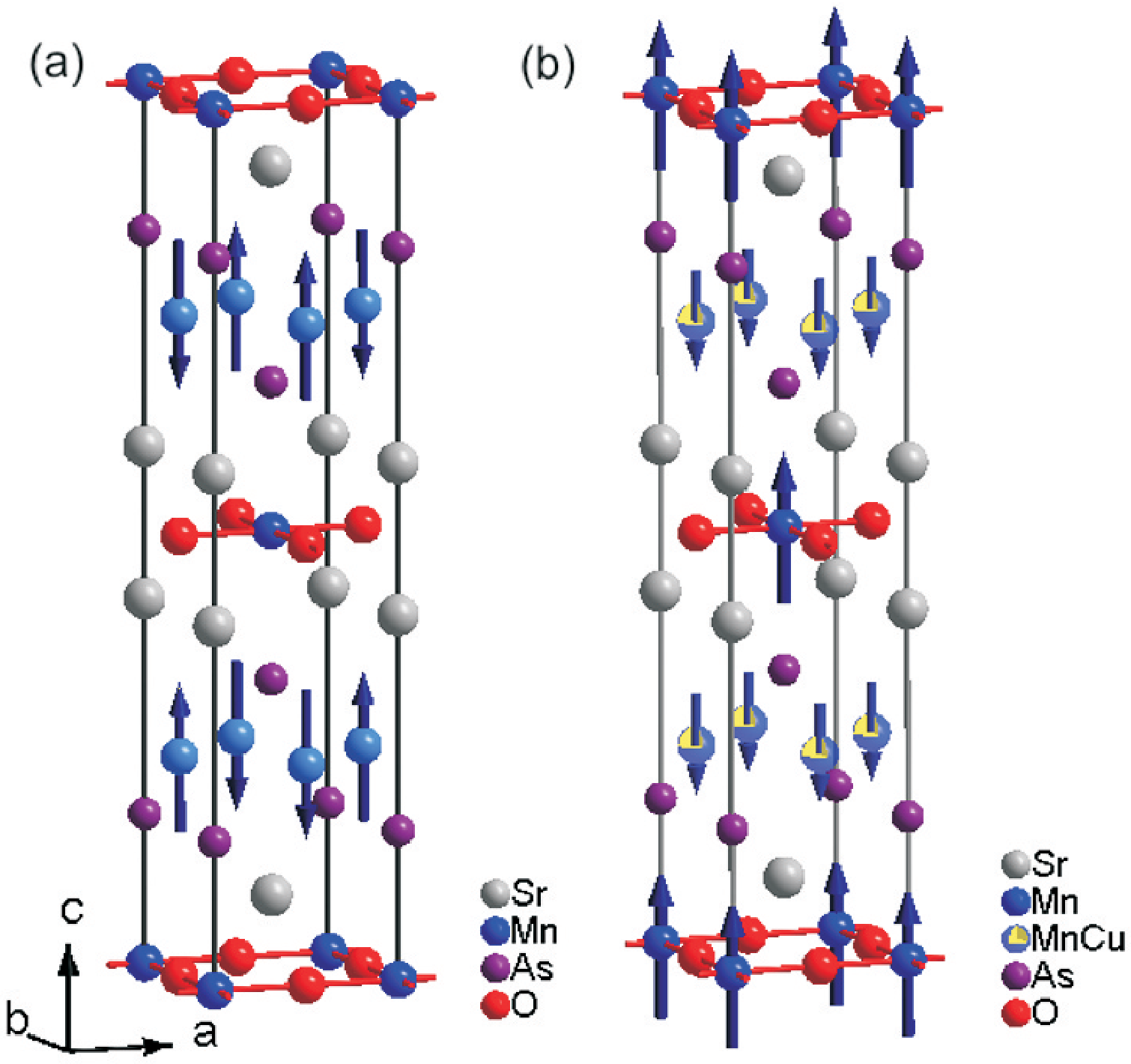}
\caption{\label{magstr}(Color online) (a) Magnetic structure model used in the 
refinement of Sr$_{2}$Mn$_{3}$As$_{2}$O$_{2}$, consisting of antiferromagnetic 
ordering of the Mn2 site  and (b) Ferrimagnetic structural model proposed for 
Sr$_{2}$Mn$_{2}$CuAs$_{2}$O$_{2}$, consisting of antiparallel alignment of Mn1 and Mn2 moments.}
\end{figure}

Finally, we turn to the results of the refinement of the magnetic
structure of Sr$_{2}$Mn$_{2}$CuAs$_{2}$O$_{2}$.  Figure~\ref{rietveld}(b) clearly
indicates some differences in magnetic behavior as compared to the
parent Sr$_{2}$Mn$_{3}$As$_{2}$O$_{2}$ compound.  First, the
transition temperature is reduced from approximately
340~K (Ref.~\onlinecite{brock1996}) to below 150~K\@.  Second, we see from Fig.~\ref{bragg}(b) that the intensities of both the (101) and (110) reflections increase below the magnetic transition in
contrast to what is found for Sr$_{2}$Mn$_{3}$As$_{2}$O$_{2}$ in
Fig.~\ref{bragg}(a).  The results of the refinement of the magnetic structure
at $T = 60$~K and $T = 4$~K are given in Table~\ref{22122Data} and illustrated in Fig.~\ref{magstr}(b). Consistent with the results of the bulk magnetization and magnetic susceptibility 
measurements, we find a ferrimagnetic structure with a net moment on
the MnO$_2$ planes (Mn1 site) that is oriented antiparallel to the net
moment in the Mn$_2$As$_2$ planes (Mn2 site).  Thus there is a dramatic change in the magnetic structure.  The magnetic structure has changed from a G-type antiferromagnet to an A-type ferrimagnet, where the spins within a layer are ferromagnetically aligned with each other, but adjacent ferromagnetic layers are antiferromagnetically aligned.  The ferrimagnetic state arises instead of an antiferromagnetic state because the opposing spins in adjacent layers do not have the same magnitude per formula unit.

At $T = 4$~K, the ordered moment
on the Mn1 site is 3.89(11)~$\mu_{\rm B}$/Mn1~atom (or per Mn1 site since the
Mn1 site is fully occupied by Mn), whereas the average moment per Mn2
site is reduced to 1.10(7)~$\mu_{\rm B}$/Mn1~site.  Thus the ordered moments per formula unit in the ${\rm Mn_2As_2}$ and ${\rm MnO_2}$ layers are
\bea
\mu_{\rm Mn_2As_2} &=& 2[1.10(7)] = 2.20(14)~\mu_{\rm B} \nonumber\\
&&\hspace{.5in}({\rm neutron\ diffraction})\label{neutmoments}\\
\mu_{\rm MnO_2} &=& 3.89(11)~\mu_{\rm B}. \nonumber 
\eea
These values are significantly larger in magnitude than estimated in Eqs.~(\ref{nonzerochi0}) and~(\ref{zerochi0}) from a fit to the $\chi(T)$, but this disagreement is not too surprising in view of the crude model used to fit the susceptibility data to obtain the latter two sets of values.  The net ferrimagnetic moment at $T = 4$~K from our neutron diffraction structure refinement is $\mu_{\rm sat} = 3.89(11) - 2.20(14) = 1.7(3)~\mu_{\rm B}$/f.u., which is reasonably close to the saturation moment $\mu_{\rm sat} = 2.0~\mu_{\rm B}$/f.u.\ extrapolated to zero field from the $M(H)$ measurements at 1.8~K in Fig.~\ref{MH1}. Similar to Sr$_{2}$Mn$_{3}$As$_{2}$O$_{2}$, Sr$_{2}$Mn$_{2}$CuAs$_{2}$O$_{2}$ does not exhibit any clear additional three-dimensional long-range magnetic ordering transitions below the initial ordering temperature.

\section{\label{SecDiscuss} Discussion}

The effective moments $\mu_{\rm eff}$ for Sr$_{2}$Zn$_{2}$MnAs$_{2}$O$_{2}$ [5.57(5)~$\mu_{\rm B}$/f.u.\ from Fig.~\ref{susc2}] and Sr$_{2}$Mn$_{3}$As$_{2}$O$_{2}$ [6.14(2)~$\mu_{\rm B}$/f.u.\ from Fig.~\ref{susc3}(a)] are close to the spin-only value of 5.92~$\mu_B$ expected for the high spin state ($S=5/2$) of Mn$^{2+}$ with $g = 2$.  In Sr$_{2}$Mn$_{3}$As$_{2}$O$_{2}$ this value is attributed to the single disordered Mn/f.u.\ below the N\'eel temperature $T_{\rm N} = 340$~K at which the other two Mn/f.u. exhibit long-range antiferromagnetic order.  The Weiss temperatures in the Curie-Weiss law are +42(2)~K for Sr$_{2}$Zn$_{2}$MnAs$_{2}$O$_{2}$ and +3(1)~K for Sr$_{2}$Mn$_{3}$As$_{2}$O$_{2}$.  The $\chi(T)$ data show a bifurcation betweeen the FC and ZFC data at about 50~K in both compounds.  The $M(H)$ data for Sr$_{2}$Zn$_{2}$MnAs$_{2}$O$_{2}$ show no evidence for ferromagnetic ordering.  The combination of these data and the $\chi(T)$ data suggest spin-glass ordering at about 50~K in both compounds arising from frustration and/or competing interactions, consistent with previous reports.\cite{brock1996a,ozawa1998,ozawa2001}  From classical Monte Carlo simulations, Enjalran and coworkers concluded that single ion anistropies are needed to explain the observed magnetic structures in this class of compounds, particularly the orthogonal ordering of the spins in adjacent layers observed\cite{brock1996} in Sr$_{2}$Mn$_{3}$Sb$_{2}$O$_{2}$.\cite{Enjalran2000}

On the other hand we find that Sr$_{2}$Mn$_{2}$CuAs$_{2}$O$_{2}$ exhibits a long-range ferrimagnetic transition as deduced from both magnetic susceptibility and neutron diffraction measurements.  The latter measurements show a change in the magnetic ground state structure from a G-type antiferromagnetic structure in Sr$_{2}$Mn$_{3}$As$_{2}$O$_{2}$, in which only the Mn1 spins in the Mn$_2$As$_2$ layers appear to undergo three-dimensional long-range ordering [although the (100) magnetic peak we observe remains unexplained], to an A-type ferrimagnetic structure in Sr$_{2}$Mn$_{2}$CuAs$_{2}$O$_{2}$ in which the Mn1 ions in the MnO$_2$ layers now also exhibit long-range magnetic ordering.  A steep increase of $\chi(T)$ and a bifurcation of the ZFC and FC $\chi(T)$ data in Fig.~\ref{susc1}(a) suggest that the Curie temperature is $T_{\rm C} \sim 80$--100~K\@.  A more accurate determination of $T_{\rm C}$ was carried out using Arrott plots which gave $T_{\rm C} = 95(1)$~K with mean-field critical exponents for the susceptibility and spontaneous magnetization.  It is surprising and unexplained why our specific heat data for this compound show no discernable anomaly at $T_{\rm C}$.

Our $\rho(T)$ data indicate that Sr$_{2}$Zn$_{2}$MnAs$_{2}$O$_{2}$ and 
Sr$_{2}$Mn$_{3}$As$_{2}$O$_{2}$ are both narrow band gap semiconductors
with activation energies $\Delta = 147(5)$~meV and 133(4)~meV, respectively. The Ba analogue Ba$_{2}$Zn$_{2}$MnAs$_{2}$O$_{2}$ is also reported to be a semiconductor with $\Delta \simeq 92.2$ meV 
(1070~K).\cite{matsushita2000}

Partial substitution of Mn by Cu in Sr$_{2}$Mn$_{3}$As$_{2}$O$_{2}$ to form the new compound Sr$_{2}$Mn$_{2}$CuAs$_{2}$O$_{2}$ leads to a drastic change in $\rho(T)$ from any of the other compounds studied and we infer the ground state of this compound is probably metallic, consistent with our low temperature specific heat data that show a sizable linear specific heat coefficent.  Single crystal resistivity measurements would be helpful to confirm this hypothesis.  From our neutron diffraction studies, the Cu is found to statistically occupy 37(1)\% of the metal sites in the FeAs-type layer together with 63(1)\% of Mn. For comparison, BaMn$_{2}$As$_{2}$, in which the FeAs-type layer is completely occupied by Mn, has an insulating ground state.\cite{singh2009, an2009}  On the other hand BaCu$_{2}$As$_{2}$, in which the FeAs-type layer is completely occupied by Cu, is metallic.\cite{djsingh2009} Thus the inferred metallic ground state of Sr$_{2}$Mn$_{2}$CuAs$_{2}$O$_{2}$ is evidently specifically due to the presence of Cu in the FeAs-type layer.  

The compounds Sr$_{2}$Zn$_{2}$MnAs$_{2}$O$_2$ and Sr$_{2}$Mn$_{3}$As$_{2}$O$_2$ also show large linear heat capacity coefficients at low temperatures that cannot arise from conduction carriers since these compounds are insulators at such temperatures.  The values are  
similar to the value of 38.7~mJ/mol~K$^{2}$ reported before for the Ba analogue Ba$_{2}$Zn$_{2}$MnAs$_{2}$O$_2$.\cite{matsushita2000}  It is interesting in this context that the pure Ba compound Ba$_{2}$Mn$_{3}$As$_{2}$O$_{2}$ has a zero linear specific heat coefficient\cite{matsushita2000} as compared with the large value found for Sr$_{2}$Mn$_{3}$As$_{2}$O$_{2}$. Furthermore a significant difference was observed between the magnetic properties of these two compounds.  The $\chi(T)$ of Ba$_{2}$Mn$_{3}$As$_{2}$O$_{2}$ shows a low-dimensional ordering at $\sim 100$~K with no bifurcation between the FC and ZFC data which would have indicated the occurrence of a spin-glass transition\cite{brock1996a} while we and others\cite{brock1996a} observed such a bifurcation at $\sim 50$~K suggesting a spin-glass or some other type of transition in Sr$_{2}$Mn$_{3}$As$_{2}$O$_{2}$. Therefore it seems that the occurrence of large linear specific heat coefficients and the occurrence of spin-glass or possibly other types of transitions are closely related in this class of materials.

A comparison of the structural and magnetic properties the four compounds (1) Sr$_{2}$Mn$_{3}$Sb$_{2}$O$_{2}$, (2)
Ba$_{2}$Zn$_{2}$MnAs$_{2}$O$_{2}$, (3)
Sr$_{2}$Mn$_{3}$As$_{2}$O$_{2}$, and (4)
Sr$_{2}$Zn$_{2}$MnAs$_{2}$O$_{2}$ has been given in Ref.~\onlinecite{ozawa2001}.
The $c$-axis lattice constant 
decreases in the listed order, and the magnetic properties of the MnO$_2$ sublattice show a systematic variation. For compounds (1) and (3), the MnAs sublattice orders above 300~K in a 
G-type antiferromagnetic structure, independent of the Mn in the MnO$_2$ sublattice. Apparently only for compound (1), the Mn in the MnO$_2$ planes orders at 65~K\@.  Compounds (2) and (3) show two-dimensional short-range ordering and compound (4)  with positive Weiss temperature $\theta$ shows no clear indication of any ordering.  These systematic differences are attributed to the increasing level of ferromagnetic correlations as the $c$-axis and interlayer separation decrease.  In Sr$_{2}$Mn$_{2}$CuAs$_{2}$O$_{2}$ the $c$ lattice constant is smaller than in any of the above four mentioned compounds.  Thus the A-type ferrimagnetic
behavior of this compound is consistent with these trends.

\section{\label{SecSummary} Summary}

We have synthesized the layered compounds
Sr$_{2}$Mn$_{2}$CuAs$_{2}$O$_{2}$, Sr$_{2}$Mn$_{3}$As$_{2}$O$_{2}$, and
Sr$_{2}$Zn$_{2}$MnAs$_{2}$O$_{2}$ and investigated their
physical properties systematically by means of x-ray and 
neutron diffraction, magnetic susceptibility,
electrical resistivity, and heat capacity measurements.
Sr$_{2}$Mn$_{3}$As$_{2}$O$_{2}$ and Sr$_{2}$Zn$_{2}$MnAs$_{2}$O$_{2}$ 
were found to be narrow band gap semiconductors with activation energies
$\Delta=$~133(4) and 147(5)~K, respectively, while the new compound
Sr$_{2}$Mn$_{2}$CuAs$_{2}$O$_{2}$ appears to have a metallic ground state.  The metallic character is evidently due to the presence of Cu in the FeAs-type layer, and is also evidenced by a sizable linear specific heat coefficent.

Our magnetization, magnetic susceptibility, and neutron diffraction measurements indicated that Sr$_{2}$Mn$_{2}$CuAs$_{2}$O$_{2}$
is a ferrimagnet with a Curie temperature $T_{\rm C}=$~95(1)~K 
whereas Sr$_{2}$Mn$_{3}$As$_{2}$O$_{2}$ and 
Sr$_{2}$Zn$_{2}$MnAs$_{2}$O$_{2}$ are evidently low-dimensional systems exhibiting spin-glass transitions at about 50~K due to competing ferromagnetic and antiferromagnetic interactions, although we do not rule out long-range order of the Mn1 spins in Sr$_{2}$Mn$_{3}$As$_{2}$O$_{2}$.  Remarkably, we find that the magnetic ground state structure changes from a G-type antiferromagnetic structure in Sr$_{2}$Mn$_{3}$As$_{2}$O$_{2}$, in which only the Mn1 spins in the Mn$_2$As$_2$ layers order long-range and nearest-neighbor spins are antiparallel, to an A-type ferrimagnetic structure in Sr$_{2}$Mn$_{2}$CuAs$_{2}$O$_{2}$ in which the Mn ions in each layer are ferromagnetically aligned, but are antiferromagnetically aligned between layers.  

It would be very interesting if a compound could be synthesized in which the Mn1 atoms in the MnO$_2$ layers of Sr$_{2}$Mn$_{3}$As$_{2}$O$_{2}$ were replaced by Cu and the Mn2 atoms in the MnAs layers replaced by Fe.  This would give a compound with alternating CuO$_2$ and Fe$_2$As$_2$ layers which are the superconducting layer elements of the cuprate and pnictide high $T_{\rm c}$ superconductors, respectively.  Although our efforts to accomplish this goal have not been successful so far as described above, this is still a promising direction for future synthesis efforts.

\begin{acknowledgments}

Work at the Ames Laboratory was supported by the Department of
Energy-Basic Energy Science under Contract No.~DE-AC02-07CH11358.  
The work at the High Flux Isotope Reactor, Oak Ridge National Laboratory (ORNL), was sponsored by the
Scientific User Facilities Division, Office of Basic Energy
Sciences, U.S. Department of Energy (U.S. DOE). ORNL is operated by
UT-Battelle, LLC for the U.S. DOE under Contract No.\ 
DE-AC05-00OR22725.

\end{acknowledgments}

\end{document}